\documentclass[aps,pre,twocolumn,longbibliography,superscriptaddress,nopacs,amsmath,amssymb,letterpaper]{revtex4-1}

\usepackage[colorlinks=true,linkcolor=blue,urlcolor=blue,citecolor=blue,anchorcolor=blue]{hyperref}
\usepackage[utf8]{inputenc}
\usepackage{calc}
\usepackage{graphicx}
\usepackage{amsmath,amssymb,amsthm}
\usepackage{txfonts}
\usepackage{bm}
\usepackage{color}
\usepackage{hyperref}
\usepackage[dvipsnames]{xcolor}
\usepackage[caption=false]{subfig}

\begin{document}

\title{Phase transition in the recoverability of network history}

\author{Jean-Gabriel Young}%
\email{jgyou@umich.edu}
\affiliation{D\'epartement de Physique, de G\'enie Physique, et d'Optique, Universit\'e Laval, Qu\'ebec, QC. Canada}
\affiliation{Centre interdisciplinaire de mod\'elisation math\'ematique de l'Universit\'e Laval, Qu\'ebec, QC. Canada}
\affiliation{Center for the Study of Complex Systems, University of Michigan, MI, USA}
\author{Guillaume St-Onge}%
\affiliation{D\'epartement de Physique, de G\'enie Physique, et d'Optique, Universit\'e Laval, Qu\'ebec, QC. Canada}
\affiliation{Centre interdisciplinaire de mod\'elisation math\'ematique de l'Universit\'e Laval, Qu\'ebec, QC. Canada}
\author{Edward Laurence}%
\affiliation{D\'epartement de Physique, de G\'enie Physique, et d'Optique, Universit\'e Laval, Qu\'ebec, QC. Canada}
\affiliation{Centre interdisciplinaire de mod\'elisation math\'ematique de l'Universit\'e Laval, Qu\'ebec, QC. Canada}
\author{Charles~Murphy}%
\affiliation{D\'epartement de Physique, de G\'enie Physique, et d'Optique, Universit\'e Laval, Qu\'ebec, QC. Canada}
\affiliation{Centre interdisciplinaire de mod\'elisation math\'ematique de l'Universit\'e Laval, Qu\'ebec, QC. Canada}
\author{Laurent H\'ebert-Dufresne}%
\affiliation{D\'epartement de Physique, de G\'enie Physique, et d'Optique, Universit\'e Laval, Qu\'ebec, QC. Canada}
\affiliation{Department of Computer Science, University of Vermont, Burlington, VT, USA}
\affiliation{Vermont Complex Systems Center, University of Vermont, Burlington, VT, USA}
\author{Patrick Desrosiers}%
\affiliation{D\'epartement de Physique, de G\'enie Physique, et d'Optique, Universit\'e Laval, Qu\'ebec, QC. Canada}
\affiliation{Centre Interdisciplinaire de Mod\'elisation Math\'ematique de l'Universit\'e Laval, Qu\'ebec, QC. Canada}
\affiliation{Centre de Recherche CERVO, Qu\'ebec (QC), Canada}

\begin{abstract}
Network growth processes can be understood as generative models of the structure and history of complex networks.
This point of view naturally leads to the problem of network archaeology: reconstructing all the past states of a network from its structure---a difficult permutation inference problem.
In this paper, we introduce a Bayesian formulation of network archaeology, with a generalization of preferential attachment as our generative mechanism.
We develop a sequential Monte Carlo algorithm to evaluate the posterior averages of this model, as well as an efficient heuristic that uncovers a history well correlated with the true one, in polynomial time.
We use these methods to identify and characterize a phase transition in the quality of the reconstructed history, when they are applied to artificial networks generated by the model itself.
Despite the existence of a no-recovery phase, we find that nontrivial inference is possible in a large portion of the parameter space as well as on empirical data.
\end{abstract}

\maketitle

Unequal distributions of resources are ubiquitous in the natural and social world \cite{hebert2016constrained}.
While inequalities abound in many contexts, their impact is particularly dramatic in complex networks, whose structures are heavily constrained in the presence of skewed distributions of resources, such as edges or cliques \cite{hebert2012structural}.
For instance, the aggregation of edges around a few hubs determines the outcome of diseases spreading in a population \cite{st2018phase}, the robustness of technological systems to targeted attacks and random failures \cite{callaway2000network}, or the spectral property of many networks \cite{castellano2017relating}.
It is therefore not surprising that much effort has been devoted to understanding how skewed distributions come about in networks.
Many of the satisfactory explanations  uncovered thus far have taken the form of constrained growth processes:
the rich-get-richer principle \cite{barabasi1999emergence}, sampling space reduction processes \cite{thurner2016Understanding}, and latent fitness models \cite{bianconi2001bose}.

A common characteristic shared by these processes is that they do not---nor are they expected to---give a perfect account of real complex systems and networks \cite{simon1957models}.
Their rules are simple and only capture the essence of the mechanisms at play, glossing over details \cite{mitchell2009complexity}.
But despite these simplifications, growth processes endure as useful models of real complex systems.
At the level of macroscopic distributions, their predictions have often been found to fit the statistics of real networks to surprising degrees of accuracy \cite{hebert2015complex}.
At the level of detailed mechanisms, they have been shown to act effectively as \emph{generative models} of complex networks \cite{leskovec2008microscopic,gomez2011modeling}, i.e., as stochastic processes that can explain the minutia of a network's growth \cite{jacobs2014unified,medo2014statistical}.
This point of view has led, for example, to powerful statistical tests that can help determine how networks evolve and change \cite{pham2016joint,overgoor2018choosing}.

The notion of growth processes as generative models is now being pushed further than ever before \cite{overgoor2018choosing}.
The burgeoning field of \emph{network archaeology} \cite{navlakha2011network}, in particular, builds upon the idea that growth processes are generative models of the \emph{history} of complex networks, able to reveal the past states of statically observed networks.
This point of view is perhaps the most clearly stated in the bioinformatics literature, 
which seeks to reconstruct ancient protein-protein interaction (PPI) networks to, e.g., improve PPI network alignment algorithms \cite{flannick2006graemlin,dutkowski2007identification} or understand how the PPI networks of organisms are shaped by evolution \cite{pinney2007reconstruction}.
Indeed, almost all algorithmic solutions to the PPI network archaeology problem 
are based on explicit models of network growth (variations on the duplication-divergence principle), and take the form of parsimonious inference frameworks \cite{pinney2007reconstruction,patro2012parsimonious,patro2013predicting}; greedy local searches informed by models \cite{navlakha2011network,li2012reconstruction,li2013maximum,gibson2009reverse}; or  maximum likelihood inference of approximative \cite{zhang2010refining},  graphical \cite{dutkowski2007identification}, and  Bayesian \cite{jasra2015bayesian} models of the networks' evolution.

Less obvious is the fact that a second body of work, rooted in information theory and computer science, also makes the statement that growth processes can generate the history of real complex networks.
This second strand of literature \cite{bubeck2017finding,racz2017basic,lugosi2018finding,shah2011rumors,mahantesh2012prediction,zhu2012uncovering,magner2017times,magner2017recovery} focuses on temporal reconstruction problems on treelike networks generated by random attachment processes \cite{drmota2009random,barabasi1999emergence}.
It has led to efficient root-finding algorithms (whose goal is to find the first node) \cite{shah2011rumors,bubeck2017finding,racz2017basic,lugosi2018finding}, and to approximative reconstruction algorithms on trees \cite{magner2017times,mahantesh2012prediction,zhu2012uncovering}.
Applying any of these algorithms to a real network amounts to assuming that growth processes---here random attachment models---are likely generative models.

The goal of this paper is to investigate classical growth processes as generative models of the histories of networks, from the point of view of Bayesian statistics and hidden Markov processes.
This investigation is made possible by recent advances in particle filtering methods as applied to temporal reconstruction from static observations \cite{jasra2015bayesian,bloem2016random,del2015sequential}.
Our contribution is threefold.
First, we give a latent variable formulation of the network archaeology problem for a generalization of the classical  preferential attachment (PA) model \cite{pham2015pafit,krapivsky2000connectivity,barabasi1999emergence,albert2000topology}.
We derive all the tools necessary to infer history using the model, including: a sampling algorithm for its posterior distribution adapted from Ref.~\cite{bloem2016random};  provably optimal estimators of the history; and  efficient heuristics well correlated with these estimators.
Second, we establish the extent to which complete history recovery is possible, and, in doing so, identify a phase transition in the quality of the inferred  histories (i.e., we find a phase where recovery is impossible, and a phase where it is achievable in large networks).
Third, we demonstrate with numerical experiments that we can extract temporal information from a statically observed network not generated by the model, including an aging model \cite{dorogovtsev2000evolution} and the phylogenetic tree of the Ebola virus.
We conclude by listing a number of important open problems.

\section{Methods for network archaeology}
\label{sec:method}

\subsection{The problem}
\label{subsec:problem}
\begin{figure}
    \centering
    \subfloat[True arrival times]{
    \centering
    \makebox[0.5\linewidth][c]{
    \includegraphics[width=0.49\linewidth, trim=0cm 1.2cm 0cm 1.1cm, clip=true]{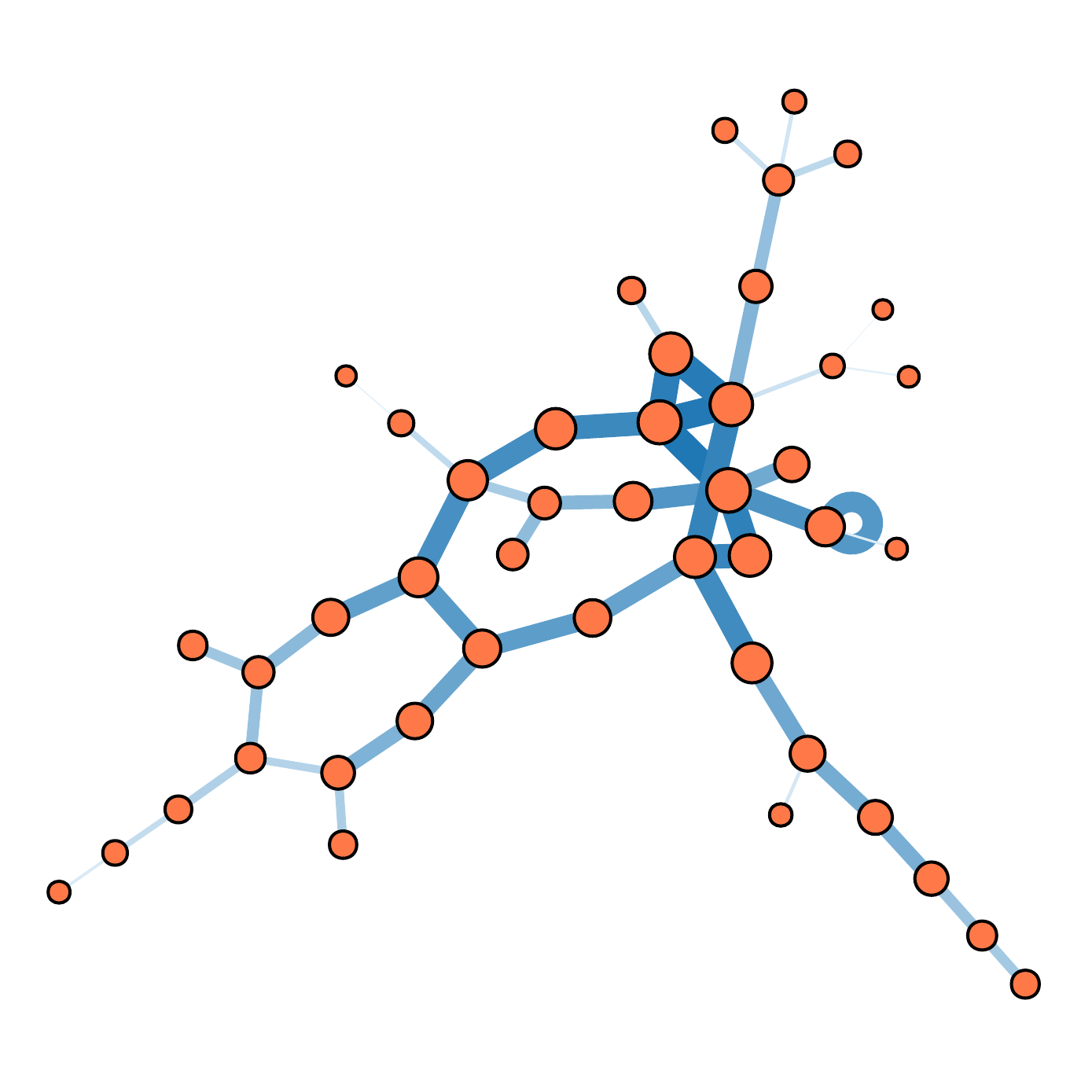}
    }
    }
    \subfloat[Posterior distribution]{
    \centering
    \makebox[0.5\linewidth][c]{
    \includegraphics[width=0.49\linewidth, trim=0cm 1.2cm 0cm 1.1cm, clip=true]{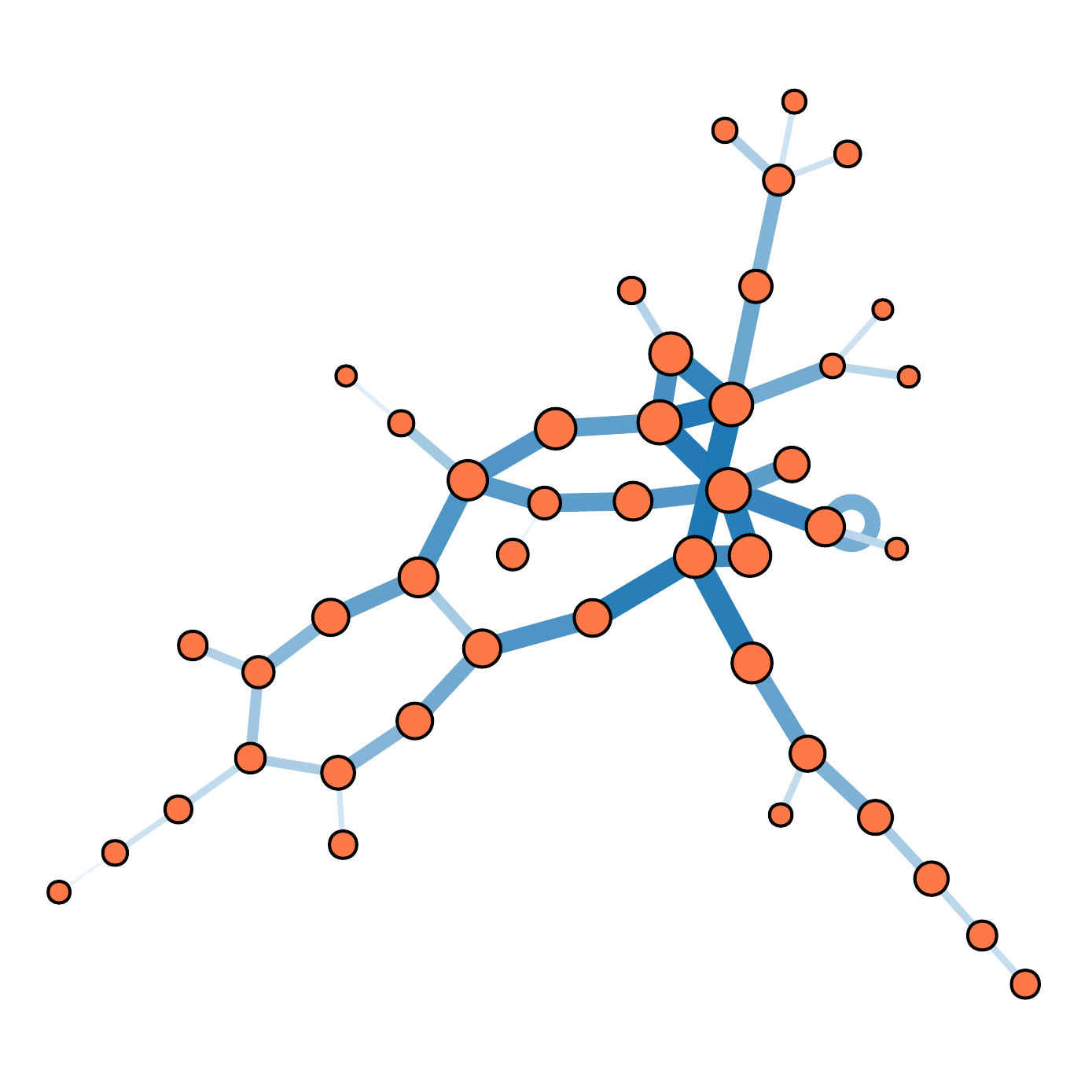}
    }
    }
    \caption{\textbf{Reconstructing the history of a growing network.}
    \textbf{(a)} Artificial network generated by our generalization of the preferential attachment model (with parameters $\gamma=-1.1, b=0.9, T=50$; see text).
    Since the network is artificial, its true history---i.e., the time of arrival of its edges in time---is known.
    The width and color of edges encode this history; older edges are drawn with thick, dark strokes, while younger edges are drawn using thin, light strokes.
    The age of nodes is encoded in their radius.
    Our goal is to infer this history as precisely as possible, using the (unlabeled) network structure as our only input.
    \textbf{(b)} Accurate reconstruction obtained with $10^5$ samples of the posterior distribution over possible histories.
    }
    \label{fig:cartoon_goal}
\end{figure}

A network $G$  generated by a growth process is, by construction, associated with a \emph{history} $X$, i.e., a series of events that explains how $G$ evolved from an initial state $G_0$.
We consider the loosely defined goal of reconstructing $X$, using the structure of $G$ and the fact that it came from a growth process as our only sources of information.
Formally, this is an estimation problem in which the history $X$ is a latent variable, determined by the structure of the network  (see Fig.~\ref{fig:cartoon_goal}).
The relationship between the network and its history is expressed using  Bayes' formula as
\begin{equation}
    P(X|G,\theta ) = \frac{P(G|X)P(X|\theta)}{P(G|\theta)}, \label{eq:bayes}
\end{equation}
where we assume, for the sake of simplicity, that the growth process parameters (denoted by the tuple $\theta$) can be estimated reliably and separately from $G$ (see Appendix.~\ref{appendix:params} where we test this assumption).

A complete specification of the probabilities appearing in  Eq.~\eqref{eq:bayes} is obtained upon choosing a growth process as our model:
This choice fixes the unconditional probability $P(X|\theta)$ of the histories, the likelihood $P(G|X)$ of the network given $X$, and the evidence $P(G|\theta)$ given as the sum of $P(G|X)P(X|\theta)$ over all histories $X$.

\subsection{Random attachment model} 
\label{subsec:model}

For the sake of concreteness, we carry out our analysis of the network archaeology problem in the context of a specific growth process.
We use a variant of the classic PA model that incorporates both a nonlinear attachment  kernel \cite{krapivsky2000connectivity} and densification events, i.e., attachment events between existing nodes \cite{pham2015pafit,krapivsky2001degree,albert2000topology,aiello2002random}.

\subsubsection{Model description}

In this model, a new undirected edge is added at each time step $t=1,2,...,T-1$, starting from an initial network $G_0$ comprising a single edge.
With probability $1-b$ the new edge connects two existing nodes, and it connects an existing node to a new node with complementary probability $b$.
Whenever an existing node $i$ needs to be selected, it is chosen randomly with probability proportional to $k_i^\gamma(t)$, where $k_i(t)$ is its degree at time $t$ (prior to any modifications to the network), and where $\gamma$ is called the exponent of the attachment kernel.
Hence the probability of choosing node $i$ when an existing node  is needed is 
\begin{equation}
    \label{eq:general_growth_function}
    u_i(\gamma) = \frac{k_i^{\gamma}(t)}{\sum_{j\in V_t} k_j^{\gamma}(t)}  =\frac{k_i^{\gamma}(t)}{Z(t;\gamma)},
\end{equation}
where $V_t$ is the set of nodes in $G$ prior to adding the new edge.

The parameter $b\in[0,1]$ controls the density, while $\gamma\in\mathbb{R}$ controls the strength of the rich-get-richer effect.
We refer to these parameters collectively with $\theta=(\gamma,b)$.
We recover the classic PA model by setting $(\gamma=1,b=1)$ \cite{price1965networks,barabasi1999emergence}; the random attachment model with $(\gamma=0,b=1)$ \cite{drmota2009random}; one of the models of Aiello, Chung and Lu with $\gamma=1,b\in[0,1]$ \cite{chung2006complex,aiello2002random}; and an undirected version of the Krapivsky-Redner-Leyvraz generalization if $\gamma$ is free to vary and $b=1$ \cite{krapivsky2000connectivity}.
The model technically generates multigraphs for any $b<1$, although numerical experiments show that the proportion of redundant edges and self-loops decreases rapidly with increasing network size,  for all $b>0$,  when $\gamma<1$.
It is thus an appropriate model of sparse multigraphs, but also a reasonable approximation of large sparse networks, with few or no redundant edges and self-loops.

\subsubsection{Posterior distribution over histories}

The posterior distribution over histories $P(X|G,\gamma,b)$ is proportional to the product of two terms; see Eq.~\eqref{eq:bayes}.

The first term, the likelihood $P(G|X)$, only weakly depends on the details of the model.
Its role is only to enforce consistency: It equals 1 if $X$ is a possible history of $G$ under the model (in which case we say that $X$ is consistent with $G$), and it equals 0 otherwise.
In \emph{any} model where growth events consist of attaching new edges or nodes to an already existing network, a consistent history $X$ is one that describes a sequence of connected subgraphs of $G$, each a subgraph of the next.
Our model is one such process, which settles the issue of calculating $P(G|X)$.

The second term, the probability of a history $P(X|\gamma,b)$, is more sensitive to the choice of model.
We notice that the growth model defined above is a Markov process, because the probabilities of growth events are defined in terms of the current state of the network and nothing else.
Hence $P(X|\gamma,b)$ can be written as a product of transition probabilities.
Denoting by $X_t$ the state of the history at time $t$, we write $P(X|\gamma,b)=\prod_{t=1}^{T-1} P(X_{t}|X_{t-1},\gamma,b)$.
Combining Eq.~\eqref{eq:general_growth_function} with the fact that a new node is involved in the growth event with probability $b$, we find that the transition probability is given by
\begin{equation}
\label{eq:model_kernel}
P(X_{t}|X_{t-1},\gamma,b) = \xi\cdot b\cdot u_{v_1}(\gamma)  + (1-\xi)\cdot (1 - b)\cdot u_{v_1}(\gamma)\cdot u_{v_2}(\gamma),
\end{equation}
where $v_1$  is the existing node and $v_2$ is the incoming node or the node chosen to close a loop.
In this equation we use $\xi=1$ to indicate that $v_2$ is new in the transition $X_{t-1}\to X_t$, and we set $\xi=0$ otherwise.

Using Eq.~\eqref{eq:bayes}, we can then write the posterior probability $P(X|G,\gamma,b)$ of a history $X$ as
\begin{equation}
    \label{eq:posterior_model}
    P(X|G,\gamma,b) = \frac{\prod_{t=1}^{T-1} P(X_{t}|X_{t-1},\gamma,b)}{P(G|\gamma,b)} \mathbb{I}[X\in \Psi(G)],
\end{equation}
where $\mathbb{I}[X\in \Psi(G)]$ is an indicator function equal to 1 if $X$ is in the set $\Psi(G)$ of histories consistent with $G$ (and 0 otherwise), and where the normalization is given by a sum over histories in $\Psi(G)$,
\begin{equation}
    \label{eq:evidence_sum}
    P(G|\gamma,b)=\sum_{X\in\Psi(G)}P(X|\gamma,b)\;.
\end{equation}

\subsection{Inference: Goal and algorithms} 
\label{subsec:inferencegoalalgo}

In the latent variable formulation of the network archaeology problem introduced in Eqs.~\eqref{eq:bayes}--\eqref{eq:posterior_model}, reconstructing the past amounts to extracting temporal information from $G$ via the posterior distribution $P(X|G,\gamma,b)$.
We need to set our goals carefully however, since not all problems of this type are solvable.
For example, the posterior distribution in Eq.~\eqref{eq:posterior_model} is heavily degenerate---and even uniform over the set of all histories consistent with $G$ \cite{magner2017recovery} for some choices of parameters $\gamma$ and $b$ (see Appendix~\ref{appendix:uniform} for details).
As a consequence, an attainable goal cannot be to find the one true history $\tilde{X}(G)$ of $G$ because this history is generally not identifiable \cite{magner2017recovery}.

To find a better inference task, we notice that, according to the model, every growth event marks the arrival of precisely one new edge (see Sec.~\ref{subsec:model}).
Consequently, we can represent histories $X$ compactly as an ordering of the edges of $G$ in discrete time $t=0,\hdots,T-1$.
And, in turn, this representation suggests a natural  inference task, which we will henceforth adopt as our inference goal: 
estimating the \emph{individual} arrival times $\tau(e)$ of the edges $e\in E(G)$ of the network.
We can hope to get good estimates in this case, because we know of a number of network properties that correlate with the age of nodes and edges in growth models \cite{adamic2000power}.

One possible estimator $\hat{\tau}(e)$ of the arrival time of edge $e$ is the posterior average:
\begin{equation}
    \label{eq:MMSE_estimator}
    \hat{\tau}(e) = \langle \tau(e)\rangle = \sum_{X} \tau_{X}(e) P(X|G,\gamma,b),
\end{equation}
where $\tau_X(e)$ denotes the arrival time of $e$ in history $X$.
It is straightforward to show that $\langle \tau(e)\rangle$ minimizes the expected mean-squared error (MSE) on $\tau(e)$, and we therefore refer to it as the MMSE estimator of the arrival time.
It turns out that this estimator also maximizes the correlation of the full set of estimates $\bigl\{\hat{\tau}(e)\bigr\}$  and the true arrival times, when $G$ really is generated by the model  (see Appendix~\ref{appendix:optimal} for a proof)---it is therefore \emph{optimal} in some sense.

Since we are working with a distribution over histories, we can do much more than simply estimate $\langle\tau(e)\rangle$.
One posterior estimate  is particularly informative: the variance on $\tau(e)$,  calculated as
\begin{align}
    \!\!\sigma^2(e) &= \langle \tau(e)^2 \rangle - \langle \tau(e)\rangle^2\notag\\
                           &= \sum_{X} \tau_{X}^2(e) P(X|G,\gamma,b) - \left(\sum_{X} \tau_{X}(e) P(X|G,\gamma,b)\right)^2\!\!.
    \label{eq:variance}
\end{align}
It can tell us how much we should trust our estimate of $\tau(e)$.
A small variance means that the (unimodal) marginal distribution $P(\tau(e)=~\!t|G,\gamma,b)$ is peaked on a few values of time $t$, i.e., that we should be pretty confident of our estimate of $\tau(e)$.
But conversely, if it is large---say in the extreme case of a uniform marginal distribution over $t=0,...,T-1$---then it means that we do not know much about $\tau(e)$ and our estimate should not be trusted.
In the applications of Sec.~\ref{sec:results}, we will quantify our uncertainty by calculating this variance alongside our estimate.
And to summarize this information in a single number, we will compute the normalized variance per edge
\begin{equation}
    \label{eq:uncertainy}
    U(G) = \frac{12}{T(T^2 - 1)}\sum_{e\in E(G)}  \sigma^2(e),
\end{equation}
where the leading factor bounds $U(G)$ in the range $[0,1]$, with $1$ corresponding to the maximal overall variance, i.e., maximal uncertainty.

We are not ready to move on to applications, however, as the computation of a complete set of MMSE estimators and the associated uncertainty score $U(G)$ is unfortunately intractable.
Explicit summation is impossible because there are far too many histories consistent with networks of even moderate size (the upper bound $|\Psi(G)|= T!$ holds, sometimes tightly).
In general, we cannot exploit some special symmetries of $G$ to evaluate the sum, since the network is an input of the problem and therefore arbitrary.
Hence, we have to resort to approximations, which we now introduce.

\subsubsection{Sequential importance sampling}
\label{subsubsec:sis}
Following the standard practice in Bayesian statistics, we use a Monte Carlo approximation to evaluate the MMSE estimators as
\begin{equation}
    \label{eq:MMSE_tau_approx}
    \hat{\tau}(e) \approx \frac{1}{n}\sum_{i=1}^n  \tau_{x_i}(e),
\end{equation}
where $x_i$ is a random history drawn from the posterior distribution $P(X|G,\gamma,b)$.
In theory, the error on the average decreases rapidly as $O(1/\sqrt{n})$, such that we can calculate $\hat{\tau}(e)$ to a good approximation quite easily.

Things are not so simple in practice, however, because it is hard to sample from the posterior distribution $P(X|G,\gamma,b)$ directly.
The consistency constraint $X\in\Psi(G)$ depends on the minutia of the structure of  $G$ and makes generating samples from $P(X|G,\gamma,b)$ a difficult endeavor.
For this reason, and following earlier work on network history sampling \cite{wiuf2006likelihood,guetz2011adaptive},  we will prefer a simple transformation of Eq.~\eqref{eq:MMSE_tau_approx} that allows for more straightforward sampling.

The transformation relies on the introduction of a second distribution  $Q(X|G)$ over the consistent histories $\Psi(G)$.
The idea is to reexpress the MMSE estimators as
\begin{align}
    \hat{\tau}(e) &=  \sum_{X}\tau_{X}(e) P(X|G,\gamma,b)\notag\\
                  & = \sum_{X\in\Psi(G)}\frac{ \tau_{X}(e) P(X|\gamma,b)}{P(G|\gamma,b)} \frac{Q(X|G)}{Q(X|G)}\notag\\
                  & = \frac{\bigl\langle \tau_X(e) \omega(X|G,\gamma,b) \bigr\rangle_{Q}}{P(G|\gamma,b)} ,
                  \label{eq:importance_sampling_definition}
\end{align}
where the average is now computed over $Q$, and where $\omega(X|G,\gamma,b) = P(X|\gamma,b) / Q(X|G)$ is called the (unnormalized) weight of history $X$.
Equation~\eqref{eq:importance_sampling_definition} is a useful reformulation of the average in that the distribution $Q(X|G)$ is now arbitrary.
Hence, in particular, we are free to choose a distribution that is easy to sample, which allows us to evaluate Eq.~(\ref{eq:importance_sampling_definition}) as 
\begin{align}
    \label{eq:importance_sampling_implementation}
    \hat{\tau}(e) \approx \frac{1}{n P(G|\gamma,b)} \sum_{i=1}^n \tau_{x_i}(e) \omega(x_i|G,\gamma,b),
\end{align}
where the set of $n$ histories $\{x_i\}$ is now drawn from the distribution $Q$, known in this context as the proposal distribution \cite{andrieu2003introduction}.
We can safely ignore the intractable normalization $P(G|\gamma,b)$ introduced by the transformation, because the estimators $\langle \tau(e) \rangle$ must satisfy the sum
\begin{align}
    \sum_{e\in E(G)}\!\!\! \langle \tau(e) \rangle &=  \!\!\!\sum_{e\in E(G)} \!\sum_{X \in \Psi(G)}\tau_X(e)P(X|G,\gamma,b) \notag\\
                                             &= \!\!\! \sum_{X \in \Psi(G)} P(X|G,\gamma,b)\!\!\!\sum_{e\in E(G)}\tau_X(e) \notag\\
                                             &= \!\!\! \sum_{X \in \Psi(G)} P(X|G,\gamma,b) \sum_{t=0}^{T-1}t  = \binom{T}{2} ,
     \label{eq:normalization_estimator}
\end{align}
where the last equality follows from the normalization of $P(X|G,\gamma,b)$.
As a result, we may compute $\hat{\tau}(e)$ up to a multiplicative constant and use Eq.~\eqref{eq:normalization_estimator} to set the global scale.

It is advantageous to choose a Markov process as the proposal distribution $Q$, i.e., one that factorizes as $Q(X|G)=\prod_{t=1}^{T-1}Q(X_t|X_{t-1},G)$, as we then obtain what is known as a sequential importance sampling (SIS) method \cite{doucet2009tutorial}.
The method is said to be ``sequential'' because all computations can now be done on the fly.
We generate $X_{t}$ from $X_{t-1}$ with $Q(X_{t}|X_{t-1},G)$, and update the sample weight as
\begin{equation}
    \label{eq:recursive_weights}
    \omega(X_{0:t}|G,\gamma,b) = \omega(X_{0:t-1}|G,\gamma,b) \frac{P(X_t|X_{t-1},\gamma,b)}{Q(X_t|X_{t-1},G)},
\end{equation}
where $X_{0:t}=\{X_0,X_1,...,X_t\}$ refers to a history, with all states included up until time $t$.
Neither the transition $X_{t-1}\to X_t$ nor the weight update equation makes use of old states.
Hence, after the weights have been updated,  we can simply discard $X_{t-1}$ and save memory---an important benefit when the network $G$ is large.

With the above considerations in mind, we propose to use a variation on snowball sampling  \cite{erickson1979some,lee2006statistical} as the proposal distribution $Q(X|G)$.
Informally, a snowball sample mimics the growth process itself, by enumerating the edges of $G$, radiating outward from a random starting point $e_0$ (the seed).
We determine the next edge $e_1$ of the history by drawing uniformly at random from the set of edges that share at least one node with $e_0$ (excluding $e_0$ itself).
The next edge after that  is picked from the set of edges that share at least one node with $e_0$ \emph{or} $e_1$ (excluding $e_0$ and $e_1$) and so on until the graph is exhausted.

We can give a more formal definition of $Q(X|G)$ by defining the \emph{boundary} $\Omega(X_t)$ of $X_t$ as the set of all edges that share at least one node with edges already appearing in $X_t$ but that are themselves not in $X_t$.
With this notation, a snowball sample is obtained by repeatedly drawing an edge from $\Omega$ uniformly at random and updating $\Omega$ accordingly.
It is then easy to see that the transition probability associated with this proposal distribution is
\begin{equation}
    \label{eq:snowball}
    Q_{\mathrm{sb}}(X_{t}|X_{t-1},G) = \bigl[|\Omega(X_t)|\bigr]^{-1},
\end{equation}
with the convention that $Q(X_0|X_{-1},G)=1/|\Omega(X_0)|=1/|E|$.

Our choice of proposal distribution is motivated by the fact that (i) it is a Markov process such that we can use SIS, (ii) it only generates histories that are consistent with $G$, and (iii) transitions $X_t\to X_{t+1}$ can be computed efficiently.
This choice leads to an overall sampling algorithm that is itself efficient.
The worst-time complexity of generating one sample is $O(|E| \times k_{\max})$, where $k_{\max}$ is the maximal degree of $G$ (due to the boundary updates).
When this maximal degree is a slowly varying function of $|E|$, as is the case for a large portion of the parameter space \cite{krapivsky2000connectivity}, snowball sampling generates samples in near linear time in the number of edges---as fast as possible.

\subsubsection{Sequential Monte Carlo algorithm} 
\label{subsubsec:smc}
The SIS method described above is not perfect, however.
The error on the true average, for example, no longer decreases as $O(1/\sqrt{n})$, because we are not sampling from the true posterior distribution anymore.
To get an intuition as to why, notice that probability mass is conserved, such that any proposal distribution $Q(X|G)$ that is not equal to the posterior itself must, by necessity, place a low probability on some histories that have a high posterior probability.
As soon as we generate one of these rare samples, the sum in Eq.~\eqref{eq:importance_sampling_implementation} becomes dominated by a single term because of its large weight $P(X|G,\gamma,b)/Q(X|G)$, which essentially washes out the contribution of other terms.

The impact of these high weight samples on our estimates can be quantified with the \emph{effective sample size} \cite{liu1996metropolized}
\begin{equation}
    \label{eq:ESS}
    \mathrm{ESS}(\{x_i\}| G,\gamma,b) := \frac{\left[\sum_{i=1}^n \omega(x_i|G,\gamma,b)\right]^2}{\sum_{i=1}^n \omega(x_i|G,\gamma,b)^2}\;.
\end{equation}
An ESS close to $n$ tells us that all samples contribute roughly equally, while an ESS close to $0$ tells us that we only have a few useful samples at hand.
It turns out that for problems with the structure of network archaeology, the ESS of a population of samples generated with a SIS algorithm will go to 0 unless $Q$ is extremely close to $P$ \cite{doucet2009tutorial}.
Hence the SIS algorithm introduced above tends to evaluate its estimators with very few effective samples.
A natural extension of SIS called the adaptive sequential Monte Carlo (SMC) algorithm is designed to address this problem \cite{doucet2009tutorial}.

In a SMC algorithm, one still generates samples using the imperfect proposal distribution $Q$, but this is now done \emph{in parallel}.
In other words, we first pick $e_0$ for a set of $n$ histories, then $e_1$ for all these histories, and so on, all the while updating the weights with Eq.~\eqref{eq:recursive_weights}.
We could do the same with the SIS algorithm, so the two algorithms are not truly different in this regard.
The defining difference between SIS and SMC comes from the way we handle these samples.
Denote the set of parallel histories evolved up until step $t$ as  $H(t)=\bigl\{X_{0:t}^{(i)}\bigr\}_{i=1,..,n}$, and the associated set of weights by $W(t)=\bigl\{\omega_i(X_{0:t}^{(i)})\bigr\}_{i=1,..,n}$.
As we evolve the parallel histories, we monitor the ESS as a function of $t$, using Eq.~\eqref{eq:ESS} and the partial weights $W(t)$.
Whenever the ESS  becomes too small and crosses a threshold $\mathrm{ESS}^*$, we perform an additional \emph{resampling} step.
The goal of this additional step is to eliminate histories that look like they will not contribute much to our estimators.
It is implemented by creating a new set $H'(t)$ of uniformly weighted histories from $H(t)$, obtained by randomly duplicating histories with probability proportional to their current weight.
Following standard practice \cite{doucet2009tutorial}, we choose $\mathrm{ESS}^*=n/2$ as the threshold that triggers a resampling step, and we implement resampling by drawing $n$ index $\bigl\{a_{i}\bigr\}_{i=1,..,n}$ from the multinomial distribution of probabilities $\tilde{W}=\{\omega_i/\sum_j \omega_j\}_{i=1,..,n}$, and setting $H'=\bigl\{X_{0:t}^{(a_i)}\bigr\}$.
One can show that this resampling step does not bias the estimators in the limit of large $n$ \cite{doucet2009tutorial}, which gives us a method to obtain unbiased estimators of $\tau(e)$ calculated with a high ESS.

It is important to realize that while the resampling step increases the ESS by design, it does so at a cost.
A history that evolved from an unlikely starting point or in an unlikely direction has a significant probability of getting overwritten.
Erasing this history can  be a ``mistake'' in that it could  eventually evolve to a high weight state which we never get to see.
We can never make such a mistake when the number of samples $n$ is extremely large, as a few low weight histories will survive repeated resampling.
But for finite $n$, this effect---known as path degeneracy \cite{doucet2009tutorial}---can actually lead to poor inference results in practice.
We somewhat mitigate path degeneracy by resampling only when the ESS becomes small, as opposed to at every step as is done  recently in Ref.~\cite{bloem2016random}.
Even then, numerical experiments suggest that the downsides of path degeneracy outweigh the benefit of an increased ESS on loopy networks and very heterogeneous trees  (see Supplementary Information).
As a result, we use resampling only when $b=1$ and $\gamma<1$. 

This resampling technique completes the set of methods needed to draw samples from the distribution over histories.
The derivation is complex, but the end result is straightforward.
To summarize, we (i) initialize $n\gg1$ histories by setting their weights to $1$ and by drawing, for each, a seed uniformly at random from the edge set $E(G)$.
Then we (ii) evolve the $n$  histories, in parallel, using the snowball proposal distribution of Eq.~\eqref{eq:snowball} to generate random moves and Eq.~\eqref{eq:recursive_weights} to update the weights.
When $b=1$ and $\gamma<1$,  we (iii.a) keep track of the ESS with Eq.~ \eqref{eq:ESS} and trigger a resampling step whenever it drops below $n/2$.
In all other cases, we (iii.b) never resample, so we do not need to calculate the ESS.
Once we have our final set of histories, we (iv) approximate $\langle \tau(e)\rangle$ for all edges using Eqs.~\eqref{eq:importance_sampling_implementation}--\eqref{eq:normalization_estimator} and our set of samples, and we calculate the uncertainty appearing in Eq.~\eqref{eq:uncertainy} in the same way.
Our reference implementation of this method is freely available online \cite{jgyou_code}.

\begin{figure*}
\centering
\includegraphics[width=0.7\linewidth]{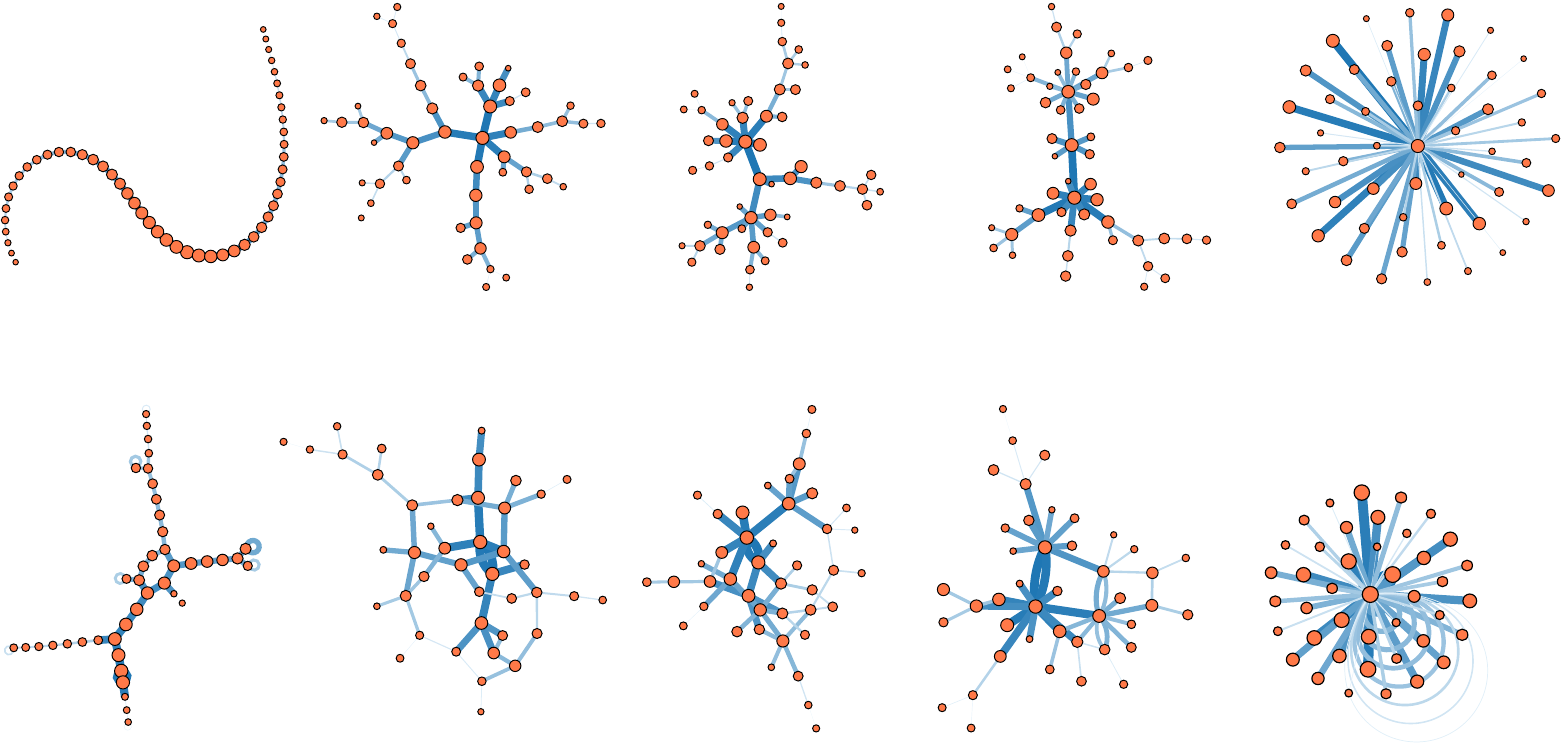}
\caption[Network zoo]
{
\textbf{Network zoo}. Examples of networks generated by the process with $b=1$ (top row) and $b=0.75$ (bottom row), and $\gamma\in\{-10,-1,0,1,10\}$ (from left to right).
The width and color of edges encode this history; older edges are drawn with thick, dark strokes, while younger edges are drawn using thin, light strokes.
The age of nodes is encoded in their radius.    
}
\label{fig:zoo}
\end{figure*}

\subsubsection{Structural estimators}
\label{subsubsec:structural_est}
We put the SIS and SMC sampling algorithms to the test in Sec.~\ref{sec:results}, but before we do, we introduce two last---much simpler---estimation algorithms as baselines.
These algorithms only rely on the structure of $G$ to estimate the arrival times $\tau(e)$, instead of an explicit knowledge of the posterior distribution $P(X|G,\gamma,b)$.
They follow the same overall pattern: 
We rank the edges in descending order, based on some network property $\mathcal{P}$ that is known to positively correlate with the age of edges, and we output these ranks as our estimated arrival times.
Whenever the edges of a subset $S\subseteq E$ are indistinguishable according to property $\mathcal{P}$, we cannot order them reliably with $\mathcal{P}$, so we instead give them the same rank $\lambda(S)$.
To set $\lambda(S)$, we require that the average time of arrival $\langle \tau \rangle$ be preserved; this constraint forces $\lambda(S)= t+(m+1)/2$, where $m=|S|$.

Our first structural estimator is based on the observation that the nodes that arrive earlier in a growth process have, on average, a larger degree \cite{adamic2000power,barabasi1999emergence}.
This result is a consequence of the fact that older nodes have many more opportunities to acquire new neighbors as the growth process unfolds than nodes that arrived at the very end.
Because we want to order edges and not nodes, we use the degree of nodes to induce a ranking of edges as follows.
We define ($k_e^{\mathrm{low}},k_e^{\mathrm{high}})$ as the degree of the nodes connected by edge $e$, with $k_e^{\mathrm{low}} \leq k_e^{\mathrm{high}}$.
We rank edges in descending order of $k_e^{\mathrm{high}}$, and break ties with $k_e^{\mathrm{low}}$.
The idea behind this strategy is that an edge connected to at least one high-degree node is likely to be older than an edge connected to two nodes of a lower degree.

We also know that the model generates networks that nucleate from a core, such that their central nodes will tend to be older.
Hence, our second structural estimator makes use of a centrality measure to order edges.
We use a recursive peeling method known as the onion decomposition (OD) \cite{hebert2016multi}, where we create a sequence of nested subnetworks by repeatedly removing the nodes of the lowest degree.
At step $t$ of this peeling process, all nodes with degree $k_{\min}$, the current lowest degree, are removed simultaneously and assigned a layer number.
We turn these numbers into the time of arrival of \emph{nodes} by assuming that nodes in the outermost layers appeared last.
A simple modification allows the algorithm to order edges: An edge is assigned to a class as soon as one of its nodes is peeled away.
All edges removed in the same pass are declared as tied.
We note that OD is closely related to the peeling method introduced in Ref.~\cite{magner2017times} to tackle the archaeology problem in the special case ($\gamma=1$, $b=1$), although the method of Ref.~\cite{magner2017times} removes the lowest-degree nodes without batching, which leads to a slightly different ordering.

\section{Results}
\label{sec:results}

\subsection{Inference on artificial networks}
To calibrate the methods and understand the conditions under which they perform well, we first apply our algorithms to networks drawn from the generative model itself.
In this situation, we know the ground truth $\tilde{X}$, and therefore the true arrival times $\tau_{\tilde{X}}(e)$ of all the edges.
As a result, we can compute the quality of our estimates $\bigl\{\hat{\tau}(e)\bigr\}$ with the Pearson product-moment correlation as
\begin{align}
    \rho = \frac{\sum\limits_{e\in E(G)} \bigl(\hat{\tau}(e) - \langle \tau \rangle\bigr)\bigl(\tau_{\tilde{X}}(e) - \langle \tau \rangle\bigr)}{\sqrt{\sum\limits_{e\in E(G)} \bigl(\hat{\tau}(e) - \langle \tau \rangle\bigr)^2}\sqrt{\sum\limits_{e\in E(G)} \bigl(\tau_{\tilde{X}}(e) - \langle \tau \rangle\bigr)^2}},
    \label{eq:pearson}
\end{align}
where $\langle \tau \rangle = (T-1)/2$ is the average arrival time, which is fixed by the choice of timescale.
This correlation takes values in $[-1,1]$, where $|\rho|=1$ indicates a perfect recovery up to a time reversal, and where $|\rho|=0$ indicates that no information is extracted from the graph at all.
It is not affected by an arbitrary linear transformation of the timescales, it penalizes spurious ordering of tied events, and it is robust to small perturbations of the estimators
\footnote{
    Even though we are comparing ``rankings'' of sorts, the Pearson correlation is better suited to our use-case than a rank correlation like Spearman's.
    The main reason is that the Monte Carlo approximations of the averages $\{\langle\tau(e)\rangle\}$ are invariably noisy, and the Pearson correlation is robust to these small perturbations, whereas a rank correlation penalizes them heavily. 
    Hence the Pearson correlation is a better choice.
    We further note that in the special case where $\hat{\tau}(e)$ takes on discrete values, the Pearson correlation of Eq.~\eqref{eq:pearson} is actually equivalent to Spearman's rank correlation, meaning that we are in effect using a rank correlation for our structural estimators.}. 

\subsubsection{Inference on artificial trees}
We consider the regime $b=1$, i.e., the regime where the model generates trees,  for our first set of experiments.
The generative model is well understood in this case, which will help us interpret the inference results more readily.
In Ref.~\cite{krapivsky2000connectivity}, it is shown that different values of $\gamma\in\mathbb{R}$ correspond to different phases that are characterized by different degree distributions (see also Fig.~\ref{fig:zoo}, upper row).
In the limit $\gamma\to-\infty$, the model generates long paths, where every node has degree 2 except for the two end nodes, of degree 1.
For all negative-values of $\gamma$,  the model favors homogeneous degrees.
When $\gamma=0$, the degree distribution is geometric, of mean 2 (since we recover the uniform attachment model \cite{drmota2009random}).
In the interval $0 < \gamma < 1$, the degree distribution takes the form of a stretched exponential, with an asymptotic behavior fixed by $\gamma$.
At precisely $\gamma = 1$, the attachment kernel becomes linear and the networks are scale-free: The degree distribution follows a power law of exponent $-3$ \cite{barabasi1999emergence}.
In the interval $1<\gamma<2$, the networks \emph{condensate} in a rapid succession of connectivity transitions at  $\gamma_m=(m+1)/m$ for $m\in\mathbb{N}^*$.
When $\gamma>\gamma_m$, the number of nodes of degree greater than $m$ becomes finite.
As a result, an extensive fraction of the edges aggregates around a single node---the condensate---and this fraction grows with increasing $\gamma$ \cite{oliveira2005connectivity}.
The condensation is complete at $\gamma=2$, where the model enters a winner-takes-all scenario characterized by a central node that monopolizes nearly all the edges.

\begin{figure}
\centering
\includegraphics[width=0.9\linewidth, trim=0cm 0.7cm 0cm 0cm, clip=true]{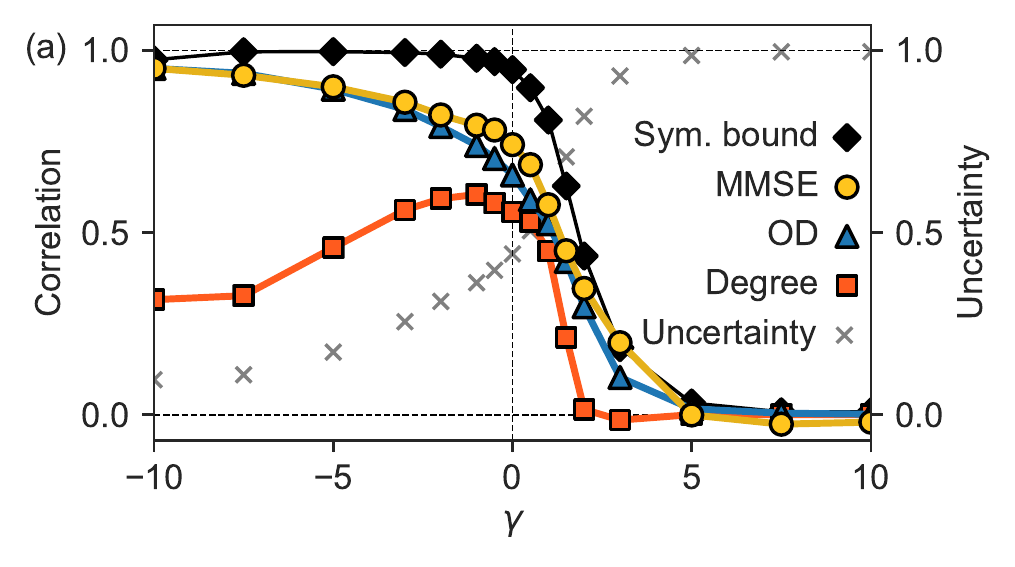}\\
\includegraphics[width=0.9\linewidth, trim=0cm 0.5cm 0cm 0cm]{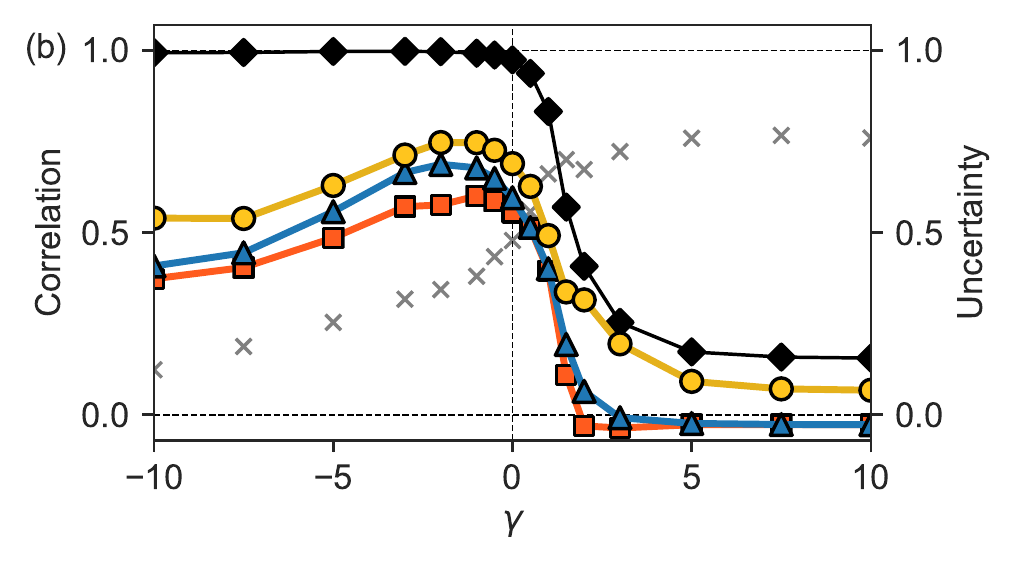}
    \caption[Effect of the rich-get-richer phenomenon on recovery]{
    \textbf{Effect of the rich-get-richer phenomenon on recovery.}
        We analyze artificial networks of $T=50$ edges,  generated with our generalization of the preferential attachment and the parameters $\gamma\in[-10,10]$, \textbf{(a)} $b=1$ (trees), and \textbf{(b)} $b=0.75$ (loopy networks).
        We plot the average correlation attained by the Monte Carlo approximation of the minimum mean-squared error (MMSE) estimators and two efficient methods based on network properties (a degree-based method and the onion decomposition \cite{hebert2016multi}).
        A loose upper bound that accounts for network symmetry is also shown, see Eq.~\eqref{eq:symbound} as well as a measure of our uncertainty of the MMSE estimators (right axis), see Eq.~\eqref{eq:uncertainy}.
        Each point is obtained by averaging the results over $m$ different network instances, where  ($b=1$) $m=40$, and ($b=0.75$) $m=250$.
        We use the true parameters $(\gamma,b)$ and $n$ Monte Carlo samples to approximate the MMSE estimators, where $n=10^5$  ($b=1$), and $n=5\times 10^6$  ($b=0.75$).
    } 
\label{fig:transitions}
\end{figure}

The average of the correlation attained by our various inference algorithms  is shown as a function of the attachment kernel $\gamma$ in Fig.~\ref{fig:transitions}a, on small networks ($T=50$).
As previously stated in Section~\ref{sec:method}, we assume that the parameters are known when we compute the MMSE estimators, and we therefore use the true values of $(\gamma, b)$ in our calculations.

We distinguish two broad regimes based on the inference results: The regime $\gamma<0$, characterized by an homogeneous distribution of degrees, and the heterogeneous regime $\gamma>0$.
The three methods behave similarly in the first regime: They first yield a relatively large correlation at $\gamma=0$, but their quality quickly plummets with growing $\gamma$, ultimately converging to a null average correlation for sufficiently large values of $\gamma$.
The MMSE estimators remain slightly superior to the OD estimators throughout, and they both outperform the degree estimators by a significant margin.
The degree estimators perform even worse in the homogeneous regime.
While the quality of the OD and MMSE estimators increases with decreasing $\gamma$, the correlation achieved by the degree estimators goes in the opposite direction and shrinks with $\gamma$.
From these results we can already draw the conclusion that the MMSE and OD estimators should be preferred on trees---MMSE for their accuracy and OD for their speed.
[It can be shown that the OD algorithm returns its final point estimates in $O(|E|\times \log|V|)$ steps \cite{hebert2016multi}, whereas a single Monte Carlo sample takes roughly as many steps to generate.]

It is natural to ask whether these results are good.
To answer this question, we also show in Fig.~\ref{fig:transitions}a two measures that can help us assess the results.

The first,  shown as gray crosses in the figure (right axis), is the uncertainty $U(G)$ defined in Eq.~\eqref{eq:uncertainy}.
It correctly increases as our ability to infer correlated histories decreases, reaching its maximal value of $1$ in the condensed phase $\gamma>2$ where inference does not seem possible.

The second, drawn in black, shows an upper bound on the average correlation as a function of $\gamma$.
This bound takes the symmetries of the generated graphs into account and shows that they can seriously hamper our ability to conduct network archaeology.

Before we describe how this bound is derived, it is helpful to understand where symmetries come from and why they matter.
Notice that some edges fulfill the same structural role in a network, say, the three edges of a triangle or the first two edges of the small network $E=\{(a,b), (a,c), (b,c), (a,d)\}$.
If we were to draw these networks twice with different layouts and the labels removed, we would not be able to tell which of  these edges is which.
In other words,  some edges can only be identified  because we have labels on the nodes.

Since labels represent an arbitrary choice \footnote{If the labels weren't arbitrary and we used the time of arrival as labels, then network archaeology would be trivial. We place ourselves in the more realistic situations where the labels are random.}, estimators cannot rely on them to make predictions.
As a consequence, structurally equivalent edges are impossible to order reliably.
Hence the more equivalent edges there are, the harder the archaeology problem becomes.

Now to actually compute the symmetry bound, we first find all the structurally equivalent edges in the generated graph $G$, using a method discussed in Appendix~\ref{appendix:orbits}.
We then construct an estimator $\tau^*(e)$ of the arrival  time of $e$ by averaging the true time of arrival $\tau_{\tilde{X}}(e)$ of all edges in its equivalence class, as
\begin{equation}
    \label{eq:symbound}
    \tau^*(e) = \frac{1}{|C(e)|}\sum_{e'\in C(e)} \tau_{\tilde{X}}(e')
\end{equation}
where $C(e)$ is the set of edges indistinguishable from $e$.
We finally compute the correlation between $\bigl\{\tau^*(e)\bigr\}$ and the ground truth, using Eq.~\eqref{eq:pearson}.
The resulting bound corresponds to the correlation we would have obtained had we known the true arrival time of edges, without the labeling of $G$.

There is no reason to believe that we can recover such a precise temporal reconstruction from $G$ alone, which means that the bound is probably loose.
That said, as shown in Fig.~\ref{fig:transitions}a, it does a good  job of explaining the maximal correlation attained in the extreme regimes---our estimators perform as well as possible when $|\gamma|\gg0$.
In the large positive-value regime, the networks condensate and symmetries upper bound the correlation at 0.
In the large negative-value regime, the networks are effectively grown as random paths, where all nodes are of degree 2 except the two end nodes, which are of degree 1 (see Fig.~\ref{fig:zoo}).
All edges are thus ordered up to a mirror symmetry around the middle, such that the equivalence classes are of size 2.
Standard concentration inequalities then tell us that the time of arrival of any edge can be identified with a variance that vanishes in the large-$T$ limit.
Near-perfect recovery is therefore trivial: Peeling the path symmetrically from both sides yields a close approximation of the arrival time of every edge.

\subsubsection{Inference on artificial loopy networks}
Figure~\ref{fig:transitions}b shows the outcome of the same experiments, in a case where loops are allowed ($b=0.75$).
The phenomenology of the generative model is not the same as in the case of trees---we do not know that there are sharp structural connectivity transitions at many values of $\gamma$, for example.
That said, the same general principle still holds (see Fig.~\ref{fig:zoo}, bottom row): Large positive-values of $\gamma$ still mean that the network condensates on a few nodes, and negative-values of $\gamma$ lead to a homogeneous distribution of degrees.

Comparing the results of Fig.~\ref{fig:transitions}b with the case of trees shown in Fig.~\ref{fig:transitions}a,  we find a number of noteworthy differences.
The most noticeable differences perhaps concern the homogeneous regime $\gamma\ll0$:
We find that near-perfect recovery is no longer possible, that the symmetry bound is much looser, and that the average uncertainty is not predictive of the reconstruction accuracy anymore.
Other important differences include an increased gap between the quality of the MMSE estimators and the structural methods (OD, degree) for all $\gamma$; and the fact that nontrivial inference remains possible deep into the condensation phase $\gamma\gg0$ with MMSE estimators.

Starting with this last difference, let us analyze the condensation phase of the generative model with loops ($b<1$), which bears a strong resemblance to the analog phase in the case of trees ($b=1$).
A typical network realization in this regime is comprised the following: many self-loops centered on the condensate, a number of parallel edges connecting high-degree nodes,  and starlike node arrangements around high-degree nodes.
In the regime $\gamma\gg0$, in particular, the typical network becomes a star with $(1-b)|E|$ self-loops on average (see bottom-right of Fig.~\ref{fig:zoo}).
These self-loops help the  degree of the condensate grow faster (incrementing its degree by 2 every time instead of 1), which leads to a more pronounced condensation for fixed $T$, $b<1$, and $\gamma>1$.
One might be tempted to conclude  that as a result, inference becomes harder as we decrease $b$, but this would be discounting the fact that self-loops carry some temporal information.
Indeed, the MMSE estimators achieve a slightly positive correlation by exploiting the difference between the self-loops and the spokes of the star in the condensation regime.
Because the growth process starts from a single edge and a geometric distribution of mean $1/(1-b)$ determines the time step at which the first self-loop is created, it is possible to obtain a positively correlated time of arrival by guessing that self-loops are slightly younger than the spokes, on average.
Neither the degree nor the OD estimators can detect this difference, and they therefore declare all edges as tied.

A different phenomenon explains the disappearance of near-perfect recovery  in the regime $\gamma\ll0$ when we set $b<1$.
The large gap between the symmetry bound and the best inference results shows that the symmetries are not at fault: One can clearly distinguish every edge, yet inference is still difficult.
We argue that the poor performance of the estimators is  instead imputable to the appearance of random long-range connections not found in other regimes.
One such edge appears when two low-degree nodes, typically located in the outermost layers of the network, are chosen as the end points of a new connection---an event that is only possible when $\gamma\ll0$ and $b<1$.
These connection in turn (i) increase the number of histories consistent with $G$, and (ii) introduce uncertainties in the ordering of large subsets of edges, for example when a long-range connection closes a long path.
The inference problem becomes harder as a result.

The correlation attained by the MMSE in the regime $\gamma\ll 0, b<1$  is probably not too far from its optimum---despite what the symmetry bound says.
The uncertainty estimates $U(G)$ are small, which tells us that the MMSE estimators are as precise as one could have hoped.
Hence a bound of a completely different nature---one that accounts for cycles---would be needed to explain the diminishing correlation as $\gamma$ goes to negative infinity.
We leave the issue of finding this bound to future work.

\subsubsection{Phase transitions in heterogeneous networks}
\label{subsubsec:phase_transition}

\begin{figure*}
    \centering
    \includegraphics[height=5.1cm, trim=0cm 0cm 0.2cm 0cm, clip=true]{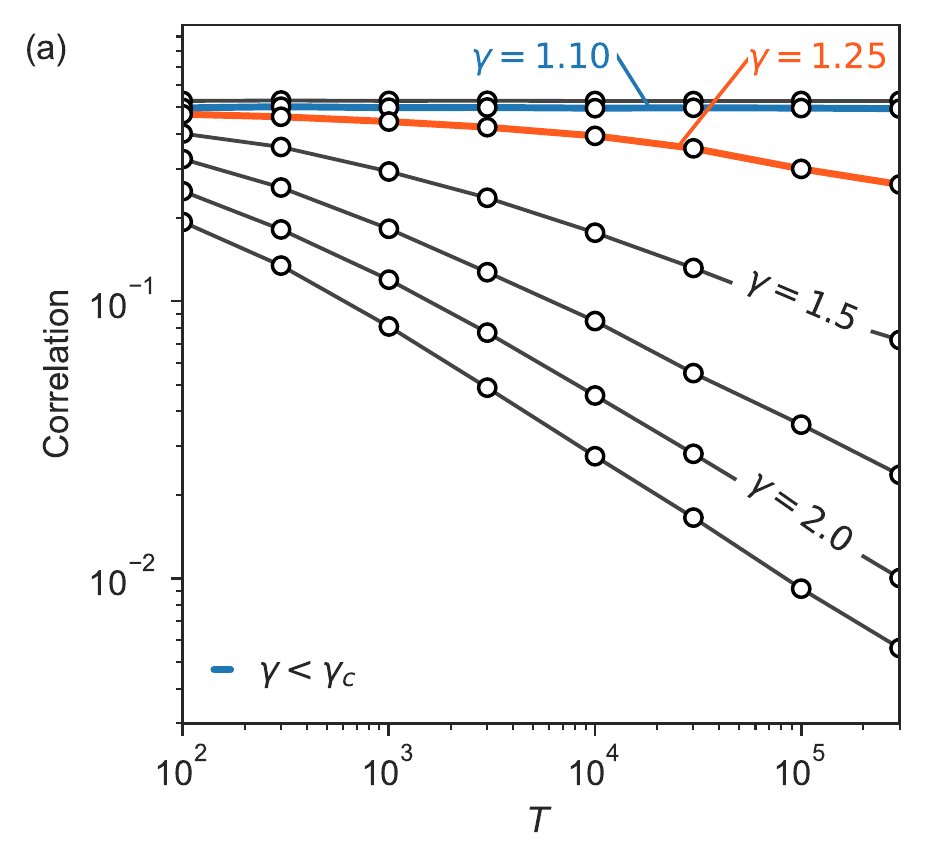}
    \includegraphics[height=5.1cm, trim=0cm 0cm 0.2cm 0cm, clip=true]{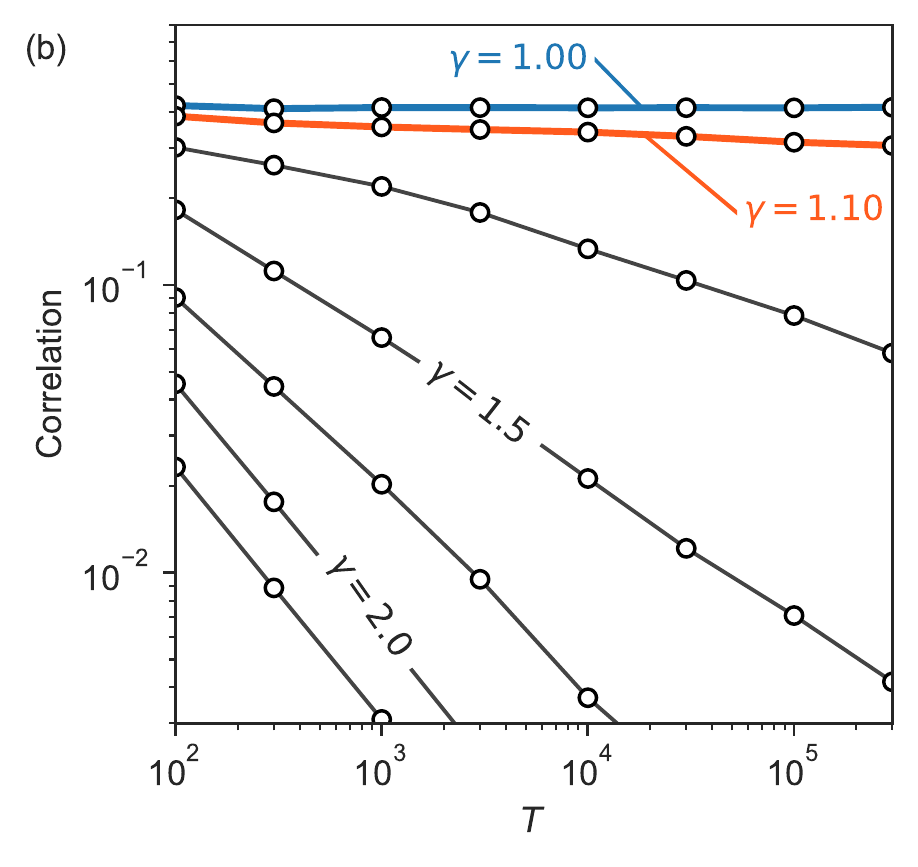}
    \includegraphics[height=5.1cm, trim=0cm 0cm 0.2cm 0cm, clip=true]{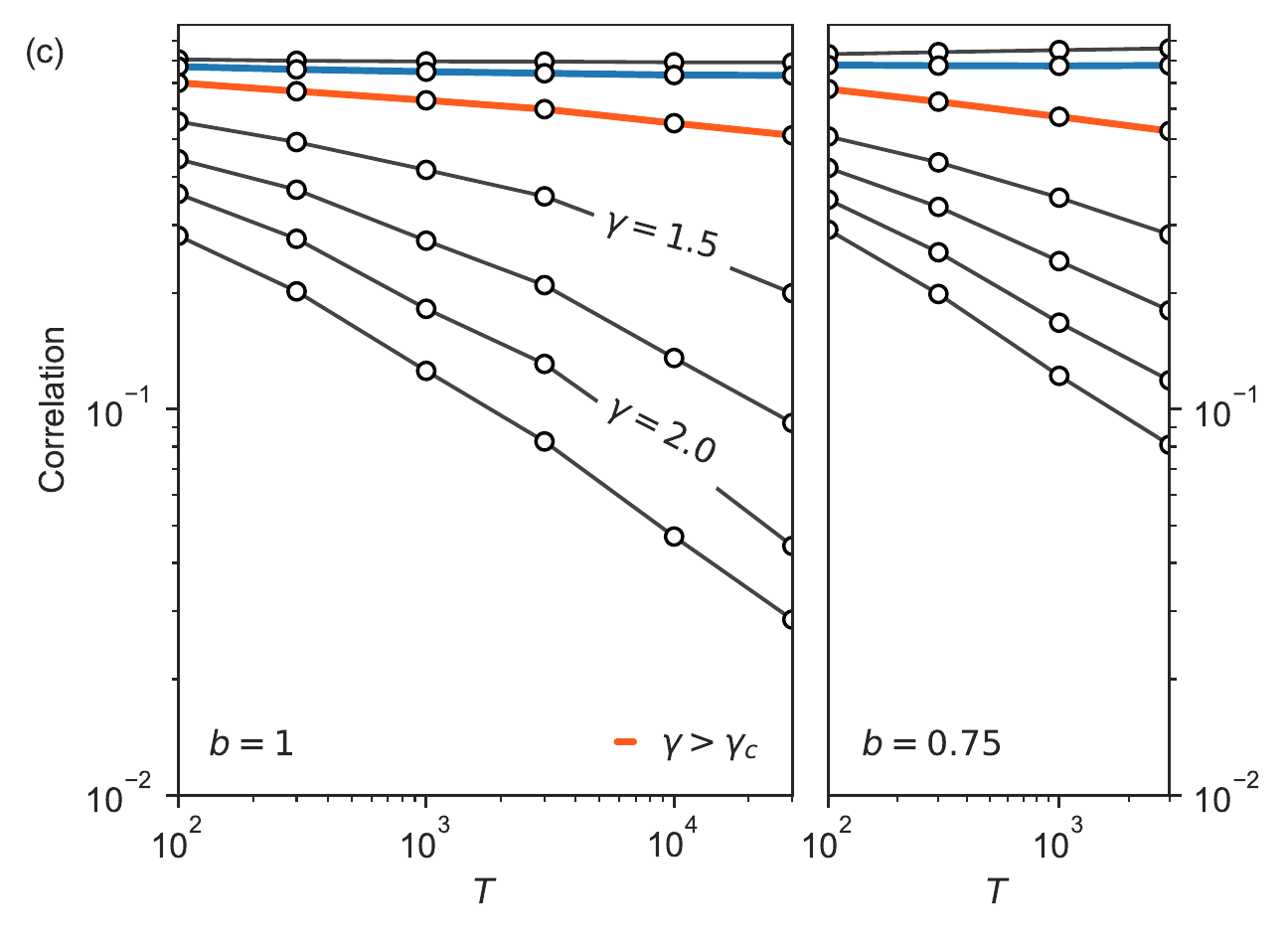}
    \caption[DFinite-size scaling analysis]{
    \textbf{Finite-size scaling analysis for the results of Fig.~\ref{fig:transitions}.} 
        The average correlation attained by OD is shown against the size $T$ of networks generated using \textbf{(a)} $b=1$ and \textbf{(b)} $b=0.75$, for $\gamma\in\{1.00,1.10,1.25,1.50,1.75,2.00,2.25\}$ (from top to bottom).
        Two scaling  curves are highlighted and annotated in each figure: The curve associated with the largest value of $\gamma$ that does not decline with $T$ (blue), and the first curve that varies with $T$ after that (orange).
        \textbf{(c)} Scaling of the symmetry bound for the same values of $\gamma$, in the case of trees (left), and dense graphs (right). The scaling is not computed for large values of $T$ due to computational costs.
        Note that there are many known structural transitions between the highlighted curves \cite{krapivsky2000connectivity}, not shown for the sake of clarity.
    }
    \label{fig:transitions_details}
\end{figure*} 

In the previous sections, we have shown that when we apply our methods to networks $G$ drawn from the generative model itself, the history encoded in the network's structure can be recovered to varying degrees of accuracy, depending on the exponent value of $\gamma$.
Focusing on the heterogeneous regime $\gamma>0$, we have seen in  Fig.~\ref{fig:transitions} that robust inference only seems possible when $\gamma$ is positive but close to 0.
These results are suggestive of a \emph{phase transition} in our reconstruction capabilities---although we cannot yet reach any conclusions because our analysis has so far been limited to small networks of $T=50$ edges, fraught with possible finite-size effects.
Therefore, to get a better numerical portrait of the dependence of the attained correlation on $\gamma$, therefore, we now turn to large networks.
Our goal is to uncover a single critical threshold $\gamma_c$ marking the onset of this transition.
We define the threshold precisely as follows:
On one side of $\gamma_c$, there exists an algorithm that returns estimators $\{\hat{\tau}(e)\}$ attaining a nonvanishing average correlation with the ground truth  in the limit of large network sizes, while there is no such method on the other side of the divide, in the \emph{no-recovery phase}.

        To find the location of $\gamma_c$, we run a finite-size scaling analysis of
        the correlation attained by different methods, on increasingly larger networks
        generated at different values of the structural transitions $\gamma_m=(m+1)/m$
        \cite{krapivsky2000connectivity}. First, we apply the OD method to these networks. Our goal is to find a
        value of $\gamma_m$ for which the correlation attained by OD is independent of
        $T$.  If we find one such $\gamma_m$, then we have evidence for a lower bound on
        $\gamma_c$ since we then know of at least one method (OD) that returns correlated
        estimates in the large-network limit.  Second, we compute the scaling of the
        symmetry bound for the same networks.  If we can find another value of $\gamma_m$
        for which the bound goes to zero as $T$ increases, then we also have an upper
        bound on $\gamma_c$ since no method can outperform the symmetry bound by
        definition. Hence these two methods combined can help us bracket $\gamma_c$.


The outcomes of these experiments are shown in Fig.~\ref{fig:transitions_details}.
We find that for most values of $\gamma>1$, the average correlation attained by the OD decreases as $T^{-\delta(\gamma)}$ with $\delta(\gamma)>0$.
If $\gamma$ is close enough to $1$, however, the average correlation becomes independent of $T$ (verified  for network sizes  up to $T\leq3\times 10^5$).
In the case of trees $(b=1)$, we find that the correlation is independent of $T$ when $\gamma=11/10$, but that is not the case for  $\gamma=5/4$, which gives us a lower bound of $\gamma_c>11/10$.
In the case of loopy graphs with $b=0.75$, we find a smaller lower bound of $\gamma_c>1$, with the correlation decreasing slowly for kernel exponents as small as $\gamma=11/10$.

Interestingly, the upper bound on $\gamma_c$ provided by the second scaling analysis (Fig.~\ref{fig:transitions_details}c) shows that the OD results are nearly optimal: When the symmetry bound decreases, the correlation attained by OD  also decreases.
Conversely, when the symmetry bound stays steady, the correlation attained by OD becomes independent of $T$.
Hence, while we have not computed the scaling analysis for all values of $\gamma$, our results suggest that $\gamma$ is bounded away from --- but close to--- $\gamma=1$  (the main structural transition).

This raises the question of where exactly the critical threshold $\gamma_c$ lies.
As we have mentioned previously, the generative model has infinitely many connectivity transitions in the case of trees, at $\gamma_m=(m+1)/m$ for $m=1,2,\hdots$ \cite{krapivsky2000connectivity}.
Our numerical results suggest that there is a \emph{single} important value of $\gamma$ that matters in the infinite size limit; as such, our parsimonious hypothesis is that one of the critical values $\gamma_m$ aligns with $\gamma_c$ when $b=1$.
Combining this simple observation with our numerical bounds leaves $m=5,...,10$ as options when $b=1$ and all values of $\gamma\in(1,1.10)$  when $b=0.75$.

\begin{figure}
    \centering
    \includegraphics[height=5.8cm, trim=0cm 0cm 0cm 0cm]{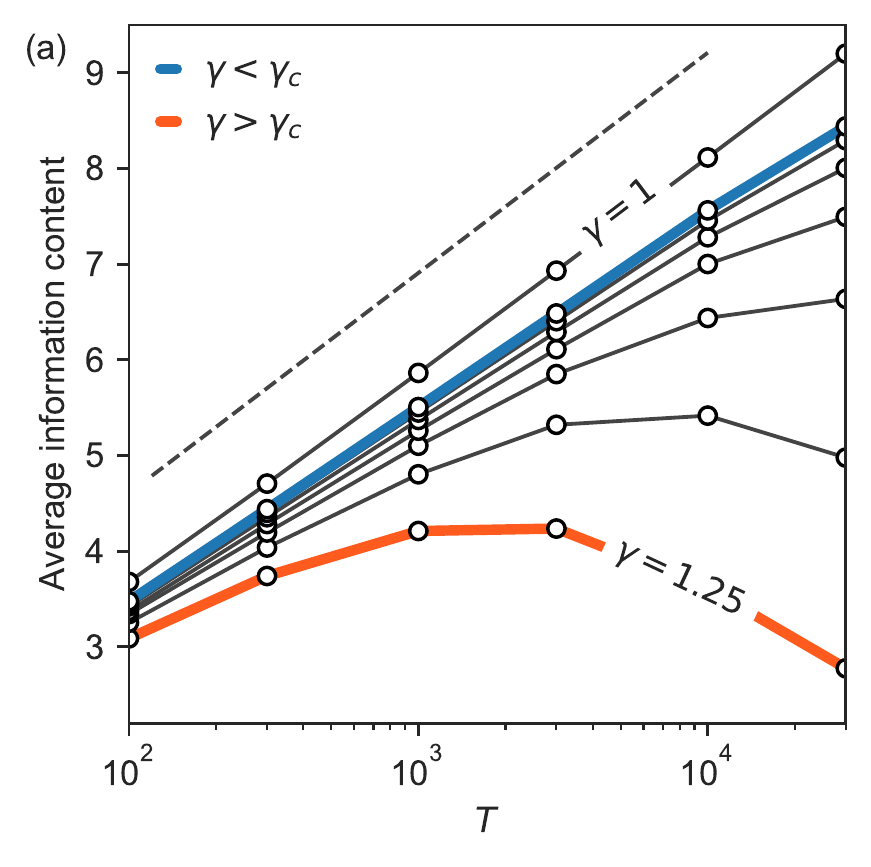}\quad
    \includegraphics[height=5.8cm, trim=0cm 0cm 0cm 0cm]{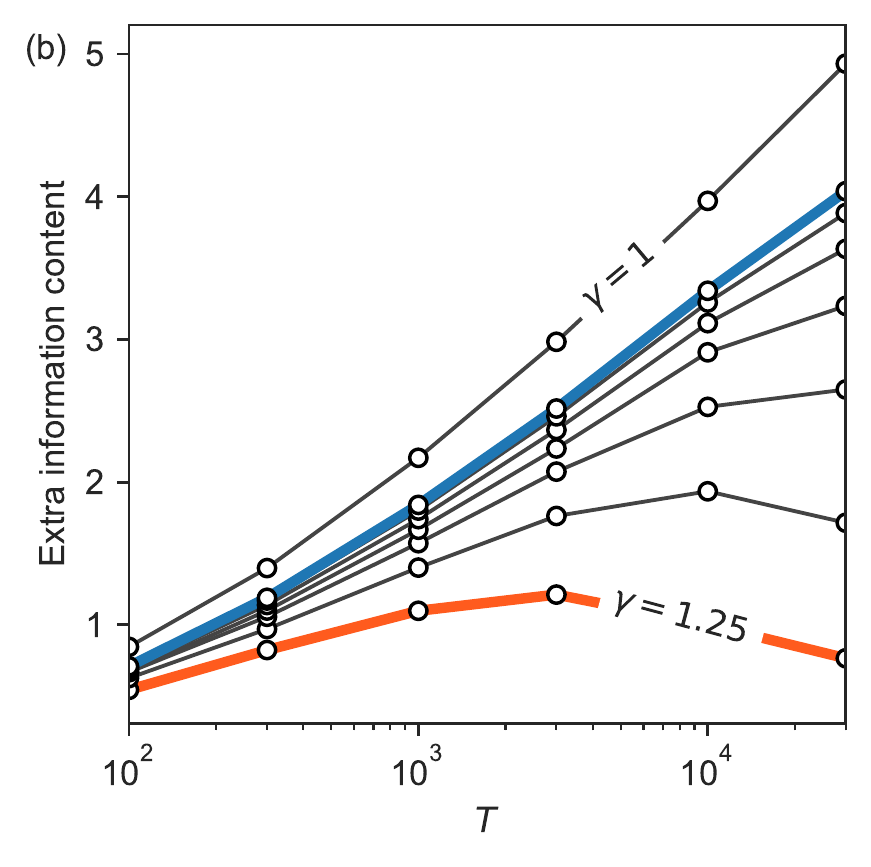}
    \caption[Average information content and recovery ]{
    \textbf{Finite size scaling of the average information content.} 
        \textbf{(a)} Average information content $\langle S(G)\rangle$ of tree networks ($b=1$), as a function of network size $T$, for exponents $\gamma\in\{\frac{5}{4},...., \frac{11}{10}, 1\}$ of the nonlinear kernel.
        Colors match what is highlighted in Fig.~\ref{fig:transitions_details}a.
        We know that these are upper and lower bounds on $\gamma_c$ due to the scaling analysis.
        The dotted line is the upper bound $\log (T)$ on the information content.
        \textbf{(b)} Information content not accounted for by the degree classes, i.e., $\langle S(G) - S_{\text{deg}}(G)\rangle$, where $S_{\text{deg}}(G)$ is the Shannon entropy of the partition obtained by classifying edges according to their nodes' degree (see text).
    }
    \label{fig:transitions_automorphism}
\end{figure} 

The appearance of a condensate from $\gamma>1$ onwards gives a nice qualitative explanation as to why there should be a phase transition in recovery quality.
When an edge attaches to the condensate, the temporal information it carries becomes inaccessible.
Furthermore, because these edges are added throughout the growth process, any estimation technique that tries to find a total ordering will conflate old and new edges in a single class.
The diminishing correlation of the estimators in the regime $\gamma>1$ is hence at least in part attributable to the presence of this condensate.

To quantify the impact of equivalent edges on the structure, we run a second scaling analysis and verify how the average \emph{information content} (IC) of the generated networks scales with network size.
The IC is, in a nutshell, an information theoretic quantity that measures the prevalence of equivalent edges\cite{rashevsky1955life}---the same edges we have used to define the symmetry bound.
It gives us a single number that summarizes the abundance but also the heterogeneity in size of the sets of equivalent edges.
It is defined as
\begin{equation}
    \label{eq:inf_content}
    S(G) = -\sum_{i=1}^{q} \frac{|C_i|}{|E|}\log \frac{|C_i|}{|E|} = \log |E| - \frac{1}{|E|} \sum_{i=1}^q|C_i|\log |C_i|,
\end{equation}
where $C_1,...,C_q$ are the $q<|E|$ sets of equivalent edges \footnote{The standard definition of $S(G)$ is in terms of nodes \cite{rashevsky1955life}; we have here opted for edges partitions since they are the basic unit of our inference.} (see Appendix~\ref{appendix:orbits} on how to find these edges).
If all the sets of equivalent edges are finite, then the information content of $G$ is of order $\log|E|$.
Conversely, if one extensive set accounts for the totality of edges---e.g., when $G$ is a star graph---then $S(G)$ is zero.
In general,  if there are $\ell$ extensive sets accounting for a nonvanishing fraction $\alpha=\alpha_1+...+\alpha_{\ell}$ of all edges, then $S(G)\approx (1-\alpha)\log|E| - \sum_i \alpha_i \log\alpha_i$.
Hence, the scaling of $S(G)$ with $|E|$ tells us how fast new sets of distinguishable edges are created  as the network grows.
What we want to verify is whether good performance correlates with the presence of many distinguishable sets of edges, i.e., a large IC.

Our results are shown in Fig.~\ref{fig:transitions_automorphism}a.
The scaling behavior of the IC confirms that there is an extensive number of equivalence classes when $\gamma=1$.
Figure~\ref{fig:transitions_automorphism}b shows the difference between the true information content and the information content obtained by assuming that the equivalence classes of edges are determined \emph{only} by the degree of the nodes at the end of the edges (a coarsening of the true equivalence classes).
Because the difference is close to zero for high values of $\gamma$, this second figure tells us that most of the equivalence classes \emph{are degree classes} in this regime.
The figure also tells us that many new equivalence classes are created as $\gamma$ approaches 1, precisely in the regime where OD does well.
Coupled with Fig.~\ref{fig:transitions_details}, Fig.~\ref{fig:transitions_automorphism} shows that an abundance of equivalent edges---specifically those of the condensate---drive the recoverability transition.

\subsection{A different task: Root-finding}
\label{subsec:rootfinding}
\begin{figure}
    \centering
    \includegraphics[width=0.8\linewidth]{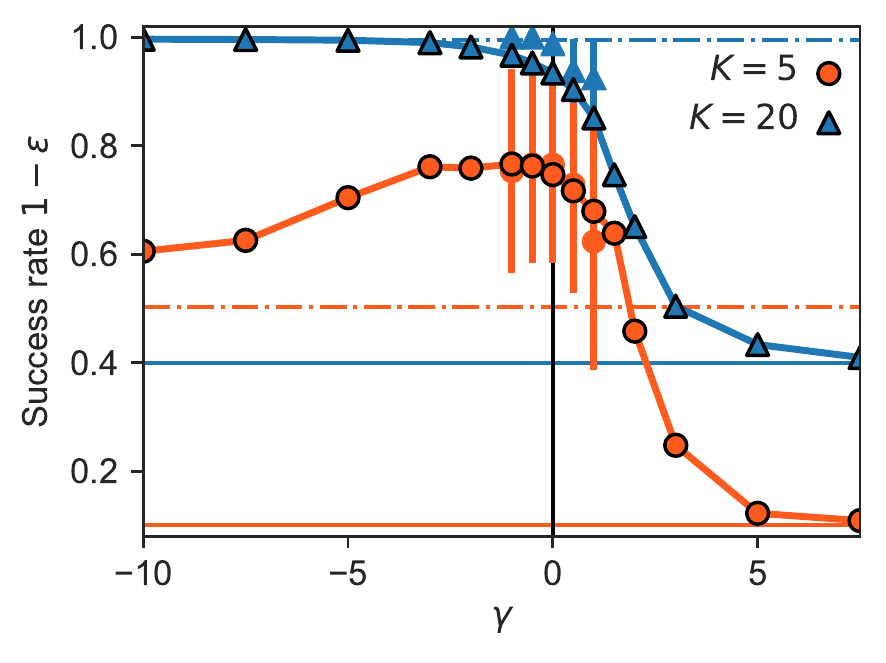}\\
    \caption[Root-finding on artificial trees]{
    \textbf{Root-finding on artificial trees}.
        Success rate of the root-finding algorithms, with sets $\mathcal{R}$ of sizes $K=5$ and $K=20$ on artificial networks of $T=50$ edges.
        The results of OD are shown with solid lines and symbols, while the sampling results are shown with symbols and error bars of 1 standard deviation.
        The horizontal solid lines show the accuracy of randomly constructed sets $\mathcal{R}$ (no information retrieval), while the horizontal dotted lines show the expected  success rate in the limit $\gamma\to-\infty$, where a simple peeling technique is optimal.
        The sampled root sets are computed for $\gamma\in\{-1,-\frac{1}{2},0,\frac{1}{2},1\}$, using $n=10^5$ samples.
    }
    \label{fig:transitions_roots}
\end{figure}

In Sec.~\ref{subsec:inferencegoalalgo}, we briefly mentioned that inferring the complete history of a network is only one of many possible problems that fits within the Bayesian formulation of network archaeology.
Any other temporal inference task that makes use of the posterior distribution $P(X|G,\gamma,b)$ can be solved with the same set of tools---like evaluating the uncertainty $U(G)$.
As a further example of the versatility of this framework, we now briefly turn to another problem: Finding the first edge (root) of $G$ \cite{shah2011rumors}.

The most comprehensive analysis of a root-finding algorithm is put forward in Ref.~\cite{bubeck2017finding}, where the goal is to find the first \emph{node} of a growing tree in the case $\gamma=0$ and $\gamma=1$.
Their strategy is to compute the number  $\varphi(v)=|\{X|\tilde{\tau}_X(v)=0\}|$ of histories rooted on $v$, and to return the $K$ nodes with the largest $\varphi(v)$.
They show that this algorithm can be employed to construct sets of constant size that contain the root with a fixed error rate $\varepsilon<1$ as $T$ goes to infinity, and that the case $\gamma=0$ is easier than the case $\gamma=1$ (smaller sets are needed to attain the same error rate $\varepsilon$).
In line with this reference and following Refs.~\cite{bubeck2017finding,lugosi2018finding,racz2017basic}, we can give a solution to the problem in terms of sets---with the crucial difference that we look for the first edge instead of the first node.

We use the marginal distribution $P(X|G,\theta)$ over time steps $P(\tau(e)=t|G,\gamma,b)$ to construct the set  $\mathcal{R}$ of likely roots.
More precisely, we use the SMC sampler to approximate the probability that an edge $e$ is the first, 
\begin{align}
    \label{eq:epoch_prob}
    P(\tau(e)=0|G,\gamma,b)=\sum_{X} \mathbb{I}\left[\tau_{X}(e) =0\right]P(X|G,\gamma,b),
\end{align}
where $\mathbb{I}[S]$ is an indicator function, which is equal to 1 if the statement $S$ is true, and equal to 0 otherwise.
We then define $\mathcal{R}$ as the set formed by the $K$ edges that have the largest posterior probability  $P(\tau(e)=0|G,\gamma,b)$, which gives us our prediction.
For comparison, we also infer the root with the much-faster onion decomposition by constructing $\mathcal{R}$ with the $K$ most central edges (with ties broken at random).

The accuracy of the resulting algorithms is shown as a function of $\gamma$ in Fig.~\ref{fig:transitions_roots}.
We distinguish, again, two main regimes:
Accurate recovery is possible in the strongly homogeneous regime $\gamma\ll0$, but the success rate diminishes with growing $\gamma$, reaching a noninformative limit in the regime $\gamma\gg 0$.
The results shows that much like full temporal reconstruction, root-finding is also negatively affected by the presence of equivalent edges.

It is worth noting that our results (Fig.~\ref{fig:transitions_roots}) put the work done in Ref.~\cite{bubeck2017trees,lugosi2018finding} in the broader context of Bayesian inference.
Their counting strategy, based on calculating the number $\varphi(v)$ of histories rooted on $v$ for example, can be formally related to our posterior inference formulation via
\begin{equation*}
    \varphi(v)\propto \!\!\!\sum_{X\in\Psi(X)}\mathbb{I}[\tilde{\tau}_X(v)=0]P(X|G,\gamma)\equiv P(\text{$v$ is first}| G, \gamma,b),
\end{equation*}
because the posterior distribution $P(X|G,\gamma,b)$ is uniform over all histories  (see Appendix \ref{appendix:uniform} for a proof) in the case studied in those works ($b=1$ and $\gamma=0,1$).
Thus the estimators proposed in Refs.~\cite{bubeck2017finding,racz2017basic,shah2011rumors} can be seen as outputting the $K$ nodes that have the largest marginal distribution at $t=0$---assuming a uniform distribution over histories.
In contrast, our method remains correct for arbitrary values of $\gamma$ (a generalization suggested in Ref.~\cite{bubeck2017finding}), but also for any choice of $b$.
It can be easily extended to seeds that are not edges but instead small subgraphs \cite{lugosi2018finding}---we simply have to change the initial distribution over states used in the SMC sampler.

\subsection{Application: Effective modeling and the phylogenetic tree of the Ebola virus}
\label{subsec:application_real}

The end goal of network archaeology is to uncover temporal information from statically observed networks not explicitly generated by growth processes.
It was recently shown  that it can sometimes be difficult to tell  growth processes apart, even when perfect temporal data are available \cite{overgoor2018choosing}.
Turning this observation on its head: there are situations where the details of the growth process do  not matter much---many mechanisms can explain the same data equally well.
One consequence of this finding is that our generalization of PA---and many more models---could find application as \emph{effective} models of the growth of real networks, without getting all the details right.
In other words, we can hope to make reasonable temporal inference even if we use an otherwise simplistic growth process as our model.

\begin{figure}
    \centering
    \includegraphics[width=\linewidth]{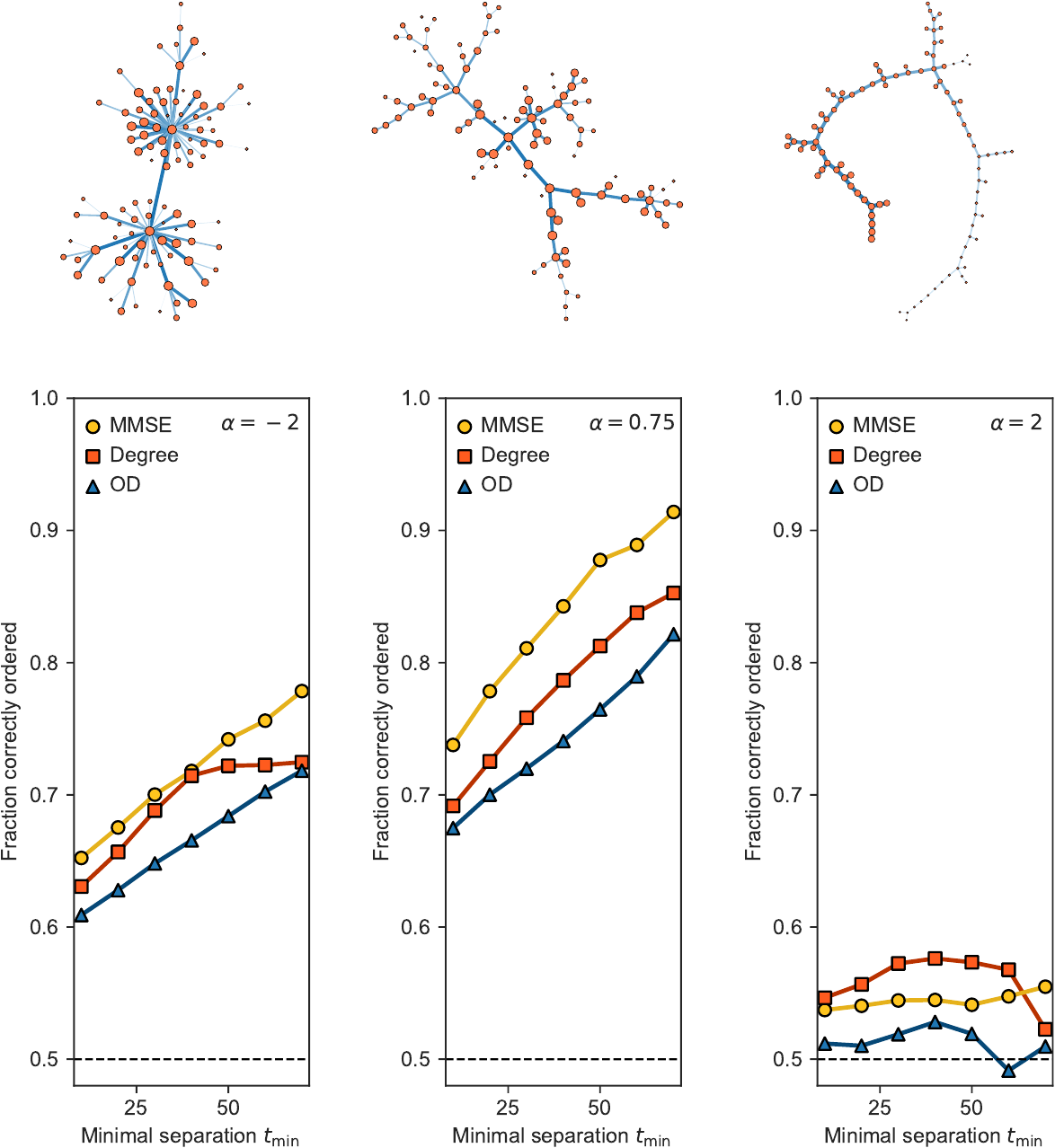}
    \caption[Application to synthetic networks grown with a model that includes memory effects]
    {
    \textbf{Application to synthetic networks grown with a model that includes memory effects.}
    \textbf{(top row)} Single network instances with various values of the memory parameters $\alpha$ \cite{dorogovtsev2000evolution}. 
    Edge thickness and node size encode age according to the ground truth (thicker and bigger equals older).
    Older nodes have a large probability of attracting new neighbors for the network shown on the left  (memory parameter $\alpha=-2$), while newer nodes are advantaged in the two other cases ($\alpha=0.75$ and $2$).
    The negative effect of memory on attachment probability is strongest for the network shown on the right, with $\alpha=2$.
    We find that these networks are best modeled with $\hat{b}=1$ and, from left to right, $\hat{\gamma}=1.76, 0.00, -1.61$ for the growth model (see Appendix~\ref{appendix:params} for details).
    The KS--statistics $D^*$ on our estimates of $\gamma$ are $D^*=0.044,0.055,0.052$, and the significance levels are $P(D>D^*)=0.47,0.45,0.39$, signaling a good fit (see  Appendix~\ref{appendix:params} for details).
    The uncertainty scores  are $U(G)=0.24, 0.39$ and $0.22$, see Eq.~\eqref{eq:uncertainy}. 
    \textbf{(bottom row)} Fraction of edge pairs  correctly ordered by the estimators, when separated by at least $t_{\min}$ time steps according to the ground truth, for the single instances of the model shown in the top row.
    Ties are broken at random for the degree and OD estimators.
    MMSE estimators are calculated from $n=200\,000$ Monte Carlo samples, using the estimated parameters.
    }
    \label{fig:aging_model}
\end{figure}

To put this hypothesis to the test, we first use our methods to reconstruct the past of an artificial model that is \emph{not} generated by our generalization of preferential attachment.
To this end, we consider a growth model that combines preferential attachment and added memory effects \cite{dorogovtsev2000evolution}.
At each time step, a new node is added to the network, and it chooses its neighbor with probability proportional to
\begin{equation}
    u_i(\alpha) = k_i(t)\  \bigl(t - \tau(v_i) + 1\bigr)^{-\alpha},
\end{equation}
where $k_i(t)$ is, again, the degree of node $v_i$ at time $t$, $\tau(v_i)$ is its time of arrival, and $\alpha\in\mathbb{R}$ is a parameter of the model.
Much like our generalization of PA, this model has a rich phenomenology, see the top row of Fig.~\ref{fig:aging_model}.
The classical preferential attachment model \cite{barabasi1999emergence} is recovered by setting $\alpha=0$.
Older nodes are chosen preferentially in the regime $\alpha<0$, whereas the newer ones are preferred in the regime $\alpha>0$.
This effect is the strongest in the regime $\alpha\geq 1$, where the network is no longer scale-free and tends to organize in long chains as $\alpha$ goes to infinity \cite{dorogovtsev2000evolution}.

Our inference results are shown in Fig.~\ref{fig:aging_model}, where we apply our methods to a few typical networks generated with this model.
Since the parameters $(\gamma,b)$ are not known in this case, we estimate them with a method detailed in Appendix~\ref{appendix:params}---this  method allows us to compute the probability $P(X|G,\gamma,b)$ when we approximate  the MMSE estimators.
For the network generated with $\alpha=-2$, we find that the best fit is given by the exponent $\hat{\gamma}=1.76$.
In other words, in this case, aging helps nodes accumulate neighbors faster than PA would, which leads to networks reminiscent of those generated by generalized PA in the superlinear regime of Ref.~\cite{krapivsky2000connectivity}.
As a result, even though our model is, strictly speaking, the wrong one, it can still give useful inference since the mechanisms behave similarly (with the difference that there is now more than one condensate node).
For the network generated with $\alpha=0.75$, we find that the best fit is  given by  the exponent  $\hat{\gamma}\approx0$.
Older nodes now fade out of memory so fast that the beneficial effect of PA  barely allows them to attract new neighbor.
This case leads to an attachment that is now more or less uniform, which again allows us to calculate useful estimates with an effective model.

The case of $\alpha=2$ is interesting---it shows that we should be careful when we interpret the results of temporal inference, in the situation where we know nothing of the growth mechanism that actually produced the network.
When $\alpha$ is large new nodes are moved out of memory immediately.
As a result, the model generates long chains similar to those generated by our model when we set $b=1$ and $\gamma\ll0$ (see Fig.~\ref{fig:aging_model}, top right), best fitted by  $\hat{\gamma}-1.61$.
A crucial difference, however, is that the oldest nodes are now concentrated on one side of the chain instead of in its center.
Our model is a poor description of this process, since it assumes that central nodes are the oldest, which leads us to estimates that are close to the random baseline (see Fig.~\ref{fig:aging_model}).

\begin{figure}
    \centering
    \includegraphics[width=0.75\linewidth, trim=0cm 3.2cm 0.5cm 3.3cm, angle=90, clip=true]{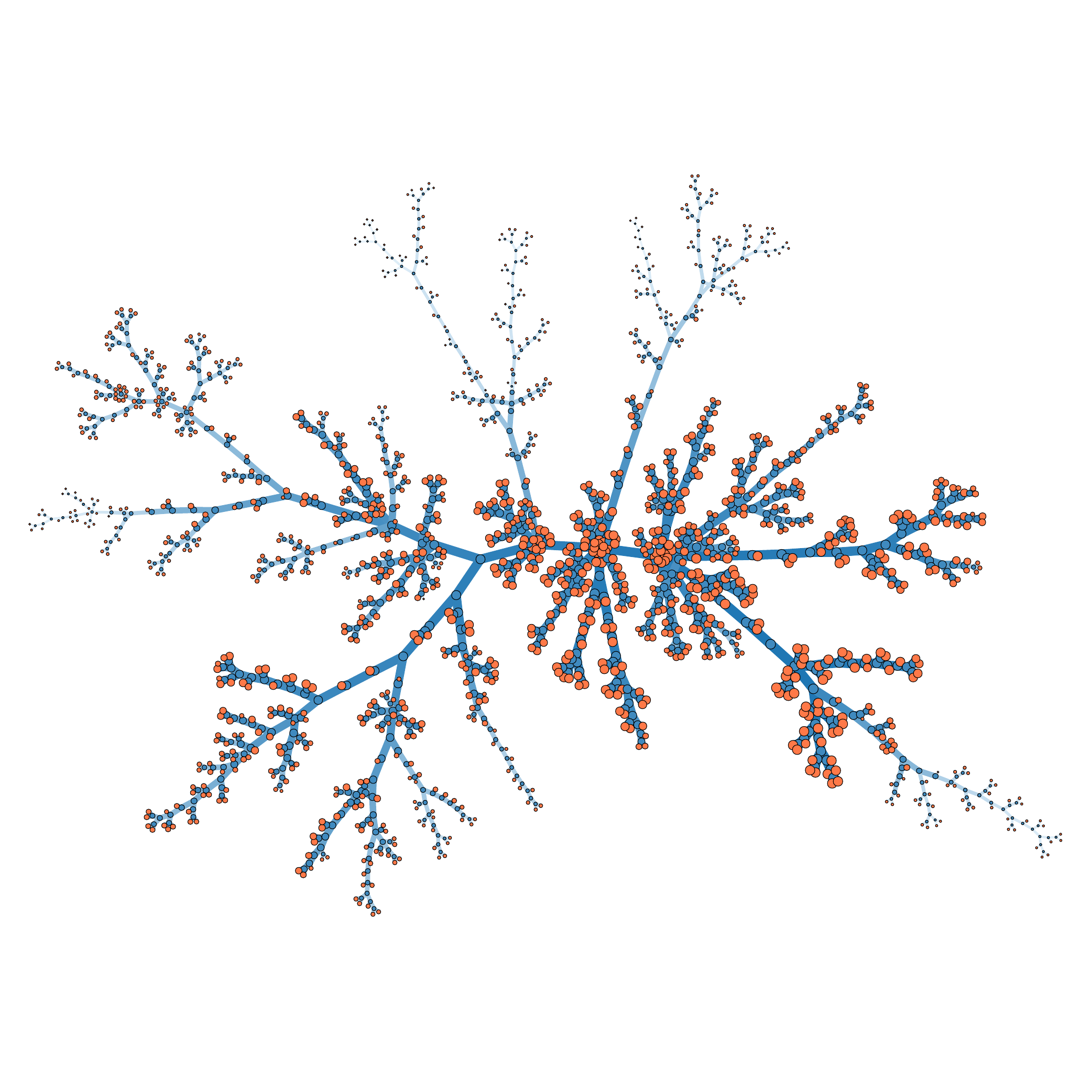} \quad
    \includegraphics[width=0.42\linewidth]{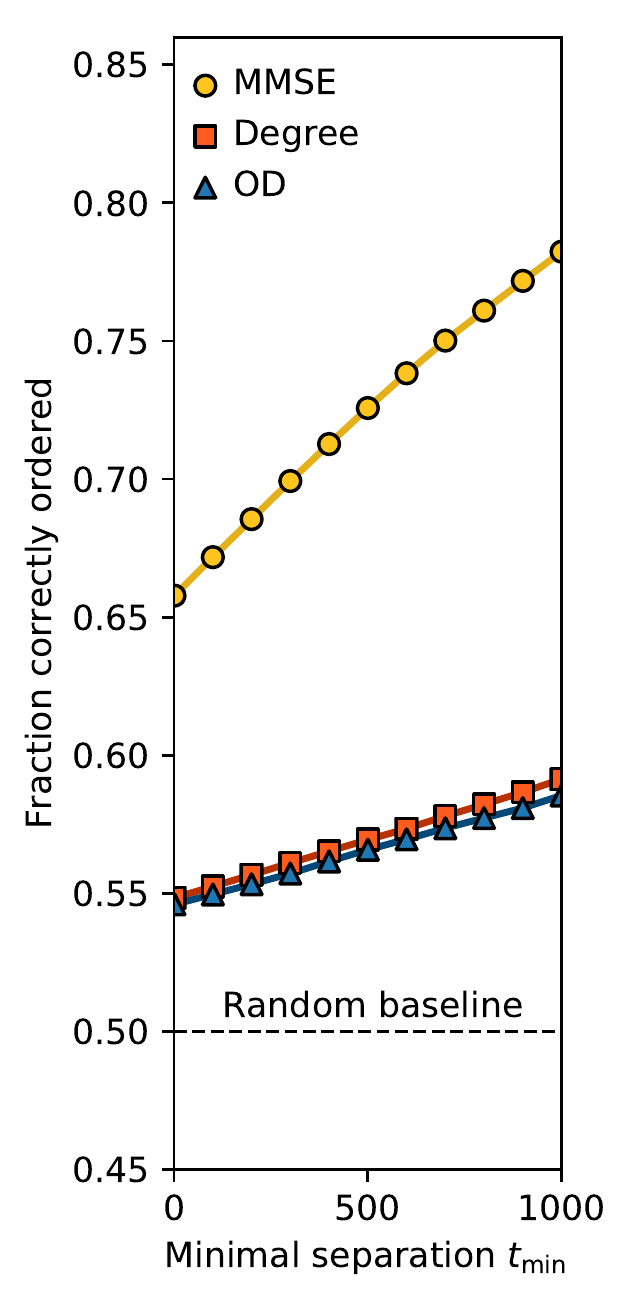}
    \caption[Application to a phylogenetic tree of Ebola strains]
    {
    \textbf{Application to the  phylogenetic tree of strains of the Ebola virus.}
    \textbf{(left)} 
    Leaves ($n=1238$, in orange) represent strains of Ebola sequenced during the 2013--2016 West African outbreak \cite{hadfield2017nextstrain}, while the remainder of the nodes represent inferred common ancestors ($n=959$, in blue), and edges are most likely mutations \cite{sagulenko2018treetime}.
    Edge thickness indicate age according to the ground truth.
    We find that this network is best modeled with $\hat{b}=1$ and $\hat{\gamma}=-0.71$, associated with a KS--statistic of $D^*=0.17$ and a significance level of $P(D>D^*)=0.53$ under the random model, signaling a good fit (see  Appendix~\ref{appendix:params} for details).
    The uncertainty score of the MMSE estimators is equal to $U(G)=0.001$, see Eq.~\eqref{eq:uncertainy}. 
    \textbf{(right)} Fraction of edge pairs  correctly ordered by the estimators, when separated by at least $t_{\min}$ time steps according to the metadata.
    Ties are broken at random for the degree and OD estimators.
    MMSE estimators are calculated with the $n=25\,000$ Monte Carlo samples.
    }
    \label{fig:ebola}
\end{figure}

With this cautionary tale in mind, we can apply network archaeology to real data---provided that we carefully choose a good growth model based on the data.
An area where network archaeology  could be useful is the study of how biological systems evolved.
Biological analyses often call for an estimation of the sequence in which the different constituents emerged, based on the currently observed biological diversity.
This information is typically encoded in phylogenetic trees that track how species evolved and diversified from common unobserved ancestors.
Much work has gone into inferring the \emph{structure} of these trees from current observations, but this structure does not necessarily tell us about the \emph{ordering} of speciation events.
Hence,  as a proof of concept, we apply our method to the inferred phylogenetic tree of the Ebola virus for the 2013--2016 West African Ebola epidemic \cite{hadfield2017nextstrain,sagulenko2018treetime}.
The relatively short duration of the epidemics and the lack of selective pressure on the virus means that Ebola underwent neutral evolution.
As a result, new lineage of strains can sprout from new and old strains alike, such that a simple attachment process is a good model of growth in this case.
The extensive coverage of the surveillance and sequencing effort for this epidemic  \cite{gire2014genomic} means that on top of the structure, we have access to temporal metadata that are a close approximation for strain emergence.
In this case, we can treat the phylogenetic tree as ground truth; our goal is to find an ordering of the emergence of all strains consistent with the metadata.

In Fig.~\ref{fig:ebola}, we show that all the inference methods recover some level of temporal information; statistical inference, however, performs much better than the others, regardless of the measure of quality used.
The naive estimators yield correlations of $\rho_{\mathrm{degree}}=0.152$ and $\rho_{\mathrm{OD}}=0.150$  with the known metadata, while we find $\rho_{\mathrm{MMSE}}=0.456$ with the sampling method (using $25\,000$ samples).
Furthermore, while all the methods resolve pairs of mutations separated by any number of time steps better than chance, the MMSE estimators outperform the other techniques.
This performance gap is in part due to the presence of equivalent edges; the OD identifies only 26 sets of distinguishable edges, while the degree estimators identify 47.
In contrast, the true MMSE estimators can order all pairs of edges that are not equivalent (there are 1588 sets of distinguishable edges), a property that is retained by the subsampled estimators.

Similar analyses are not possible for most epidemics, due to a lack of data.
For epidemics that are more  sparsely monitored than the 2013--2016 Ebola outbreak, reconstruction typically relies on models that are hard to parametrize (transmission rates, mutation rates, demographics, etc.) \cite{bouckaert2014beast}.
The quality of our results suggests that reconstructing the history of phylogenetic trees should be one of the interesting avenues for future network archaeology, especially since the method does not rely on parameters that require additional data sources.

\section{Conclusion}
In this paper, we have addressed the network archaeology problem from the point of view of Bayesian inference, with generalized preferential attachment as our generative model.
In doing so, we have shown that the equivalent edges that appear in random growing trees make inference difficult, to the point where inference becomes impossible in some regimes.
This difficulty does not mean that problem problem is generally impossible, however, since we have also shown that reconstruction is possible in both artificial and real systems.

The opportunities brought about by network archaeology are tantalizing.
In bioinformatics alone---the only field where it has found widespread use thus far \cite{jasra2015bayesian,wang2014computational}---network archaeology with the divergence-duplication models has already yielded insights into the past states of real PPI networks \cite{navlakha2011network,pinney2007reconstruction} and improved on network alignment \cite{flannick2006graemlin,dutkowski2007identification}.
Generalization to models that are relevant beyond bioinformatics will allow us to answer new questions about the past of statically observed systems, and improve network analysis techniques \cite{hajek2018recovering}.

Our paper shows an example of how to carry this analysis almost automatically.
One simply has to define a  Markov growth process;
our adaptive sequential Monte Carlo algorithm---an extension of the algorithms proposed in Refs.~\cite{bloem2016random,wiuf2006likelihood,guetz2011adaptive}---can then generate weighted histories that can be aggregated as MMSE estimators to yield optimal estimators of the history of the network.

Our analysis is, of course, far from complete, and it leaves a number of important theoretical and computational problems open.
First, while we have provided compelling evidence for the existence of a scalable inference phase and a no-recovery phase, we have yet to pinpoint the location $\gamma_c$ of the transition that separates them.
Our numerical analysis suggests that it lies at some rational value $\gamma_c=(m+1)/m$ when $b=1$, but finding the exact location will require further analytical work, perhaps relying on a detailed analysis of the expected information content of the generated graphs, for example using the counting techniques introduced in Ref.~\cite{magner2014symmetry}.
Second, we have used correlation as our notion of inference quality, but a recent analysis of the special cases $\gamma=0,1$ \cite{magner2017times} instead emphasizes trade-offs between  precision---how many elements are correctly ordered---and density---how many elements we  \emph{can} order.
It would be useful to study the full range of parameters $\gamma$ under this alternate definition of quality.
Third, the phenomenology of the observed phase transition is strikingly similar to that observed in many disorder models \cite{Decelle2011b,ricci2018typology}.
While growth models are formally out-of-equilibrium processes---and thus cannot be obviously mapped onto disorder models---it will be important to establish how the network archaeology phase transition fits within the broader family of phase transitions in Bayesian inference problems.
The tools introduced in Ref.~\cite{magner2017recovery}, further work on information content, and the concept of exchangeability \cite{crane2017edge,cai2016edge} all might offer insights into this issue.
Fourth, we have shown that nonlinear preferential attachment can sometimes act as a useful effective model of a network's growth, even when the network is definitely not generated by this model.
A systematic study of how generalizable these conclusions are could also be an interesting direction for future work.
Finally, we have used the OD algorithm \cite{hebert2016multi} in large networks because sequential Monte Carlo is, ultimately, not very scalable.
We have shown that this substitution does not work in all cases because it is based on a correlation between the true arrival times and the order of peeling of a network specific to the class of models studied.
As a result, the next step for general network archaeology should be to derive efficient approximation methods that work with general models, to allow for flexible network archaeology.
These methods will have to handle models specified as chains $P(X|\theta)$ with some arbitrary notion of consistency $P(G|X)$.
The relaxation technique of Ref.~\cite{linderman2017reparameterizing} for permutation inference comes to mind, but one could also consider the message-passing algorithm~\cite{mezard2009information} and its dynamical variant~\cite{lokhov2014inferring}.

\section*{Acknowledgment}
We thank Samuel V. Scarpino, Joshua Grochow and Alec Kirkley for helpful discussions.
This work is supported by the Fonds de recherche du Qu\'ebec-Nature et technologies (JGY, EL, PD), the Conseil de recherches en sciences naturelles et en g\'enie du Canada (GSO), the James S. McDonnell Foundation Postdoctoral Fellowship (JGY, LHD), the Santa Fe Institute (LHD), Grant No. DMS-1622390 from the National Science Foundation (LHD), and the program Sentinel North, financed by the Canada First Research Excellence Fund (JGY, EL, CM, GSO, PD).
We acknowledge the support of Calcul Qu\'ebec for help with their infrastructure.

\appendix

\section{Model parameters}
\label{appendix:params}
\subsection{Estimation}
Throughout the text, we assumed that the parameters $(\gamma,b)$ are known or can be estimated, which has allowed us to use the simpler conditional posterior $P(X|G,\gamma,b)$ (as opposed to a joint distribution over histories and parameters).
We now give a method to calculate these parameters in the cases where they are not known (such as with a real graph, see Sec.~\ref{subsec:application_real}).

Estimating $b$ is easy because the observed graph can be seen as the outcome of $|E|-1$  independent and identically distributed Bernoulli trials of success probability $b$.
Every excess edge beyond the minimum needed to ensure connectivity is seen as a failure, and the total number of edges gives the number of trials.
After adjusting for the initial conditions, we obtain the unbiased  estimator
\begin{equation}
    \hat{b}(G)=\frac{|V(G)|-2}{|E(G)|-1}.
\end{equation}

\begin{figure}
    \centering
    \includegraphics[width=0.7\linewidth]{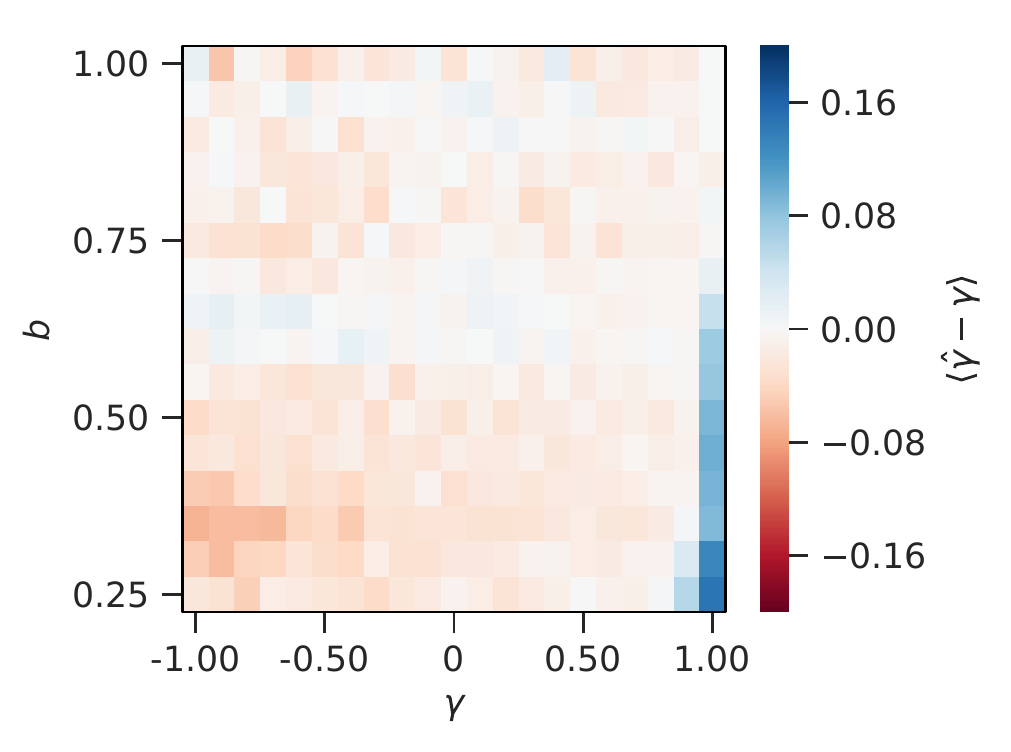}
    \caption[Consistency of the estimator of gamma in small networks]{
    \textbf{Consistency of the estimator of $\bm{\gamma}$ in small networks.}
    We show the average difference between the true value and estimated value of $\gamma$, for networks of $T=1\;000$ edges generated with parameters $\gamma\in[-1,1]$ and $b\in[0.25,1.00]$.
    This difference is averaged over 20 different network realizations at each point of the parameter space.
    We use $n=1\;000$ samples each time we evaluate the average KS--statistic.
    }
    \label{fig:esimation_consistency}
\end{figure}

The exponent $\gamma$ is, in theory, more difficult to estimate, because the degree distribution is only a sufficient statistics for $\gamma$ once it is conditioned on the arrival times \cite{gao2017asymptotic,bloem2018sampling}.
A principled estimation technique should therefore rely on a known history \cite{jeong2003measuring,gomez2011modeling} or a joint sampling mechanism for $X$ and the parameters, see Ref.~\cite{bloem2016random} for a general particle MCMC method.
But as we show in Fig.~\ref{fig:esimation_consistency}, a simple non-Bayesian heuristic---the Kolmogorov-Smirnov (KS) minimization of Ref.~\cite{clauset2009power}---yields accurate enough estimates from a single snapshot.

The KS--statistic of a pair of distributions $(P,Q)$ is given by the supremum of the difference of their cumulative distribution function (CDF), i.e.,
\begin{equation}
    D(P,Q) = \mathrm{sup}_k| f_P(k) - f_Q(k)|,
\end{equation}
where $f_P(k)$ is the CDF of $P$ at point $k$.
Given an empirical degree distribution $P(G)$ derived from an observed network $G$, we estimate $\gamma$ by minimizing the KS--statistic averaged over a set of $n$ random degree distributions $\{Q^{(i)}(\gamma)\}_{i=1,\hdots,n}$ generated by the model of parameters $(\gamma,\hat{b})$.
The minimum $D^*(G)$ can be found efficiently using Brent's method \cite{press2007numerical}, since the average KS--statistic is convex.
We use $n\gg 1$ random network instances to compute the average $D$ at each probed $\gamma$, which can be costly when $T$ is large.
Therefore, in practice, we first compute the expected degree distribution  of the model using mean-field equations, and then draw $n$ finite samples from the resulting distribution.
This approach is equivalent to---but much faster than---direct simulations.
Note that the above framework also provides a natural (non-Bayesian) notion of goodness of fit for $\gamma$ \cite{clauset2009power}.
It can assessed by generating random degree distributions with the estimated parameters $(\hat{\gamma},\hat{b})$, to which we apply the complete testing procedure.
This method provides a null distribution for $D$, which tells us whether $D^*(G)$ is an extreme value of the average KS--statistic or not.
Following the standard, we assume that if $P[D>D^*(G)]>0.1$, then the fit is good \cite{clauset2009power}.

\subsection{Sensitivity}
To show that conditioning on point estimates of the parameter does not alter temporal inference significantly, we run a sensitivity analysis in which we generate synthetic networks with some parameters $(\gamma,b)$, perturb the parameters $(\gamma,b)\to(\gamma',b')$, and then run the SMC algorithm with the perturbed parameters.
We use Fig.~\ref{fig:esimation_consistency} to select meaningful perturbation sizes: We investigate differences $\Delta_\gamma = \gamma'-\gamma$ that are much larger than the maximal error made in estimating $\hat{\gamma}$ (the error is overwhelmingly bounded to $\hat{\gamma}-\gamma\in [-0.05,0.05]$ in Fig.~\ref{fig:esimation_consistency}).
We also investigate large perturbations $\Delta_b$ to the probability $b$, but we note that large errors are unlikely since the standard deviation on the success rate of Bernoulli trials varies as $O(1/\sqrt{n})$.

\begin{figure}
    \centering
    \includegraphics[width=0.99\linewidth]{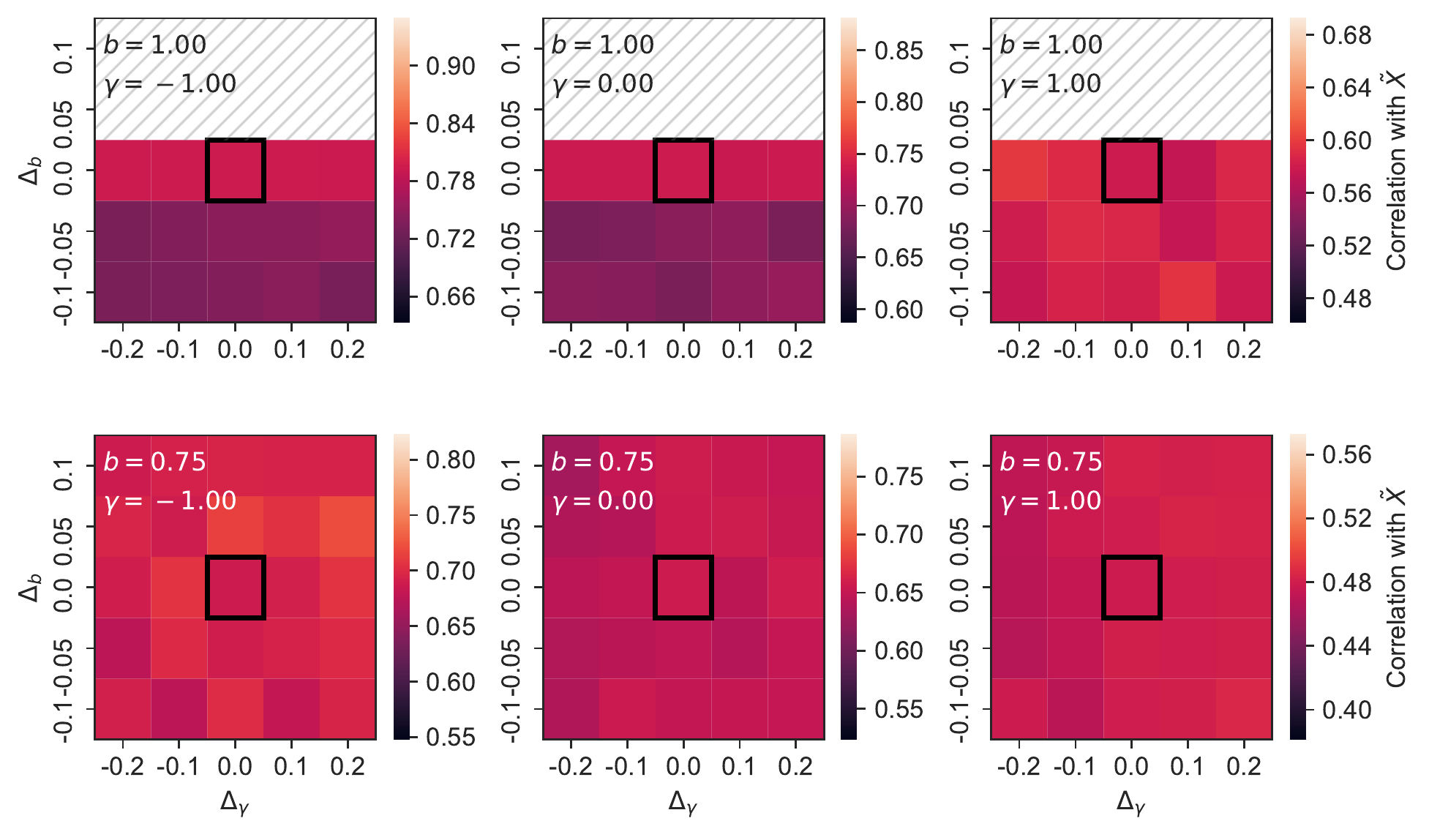}\vspace{\baselineskip}
    \includegraphics[width=0.99\linewidth]{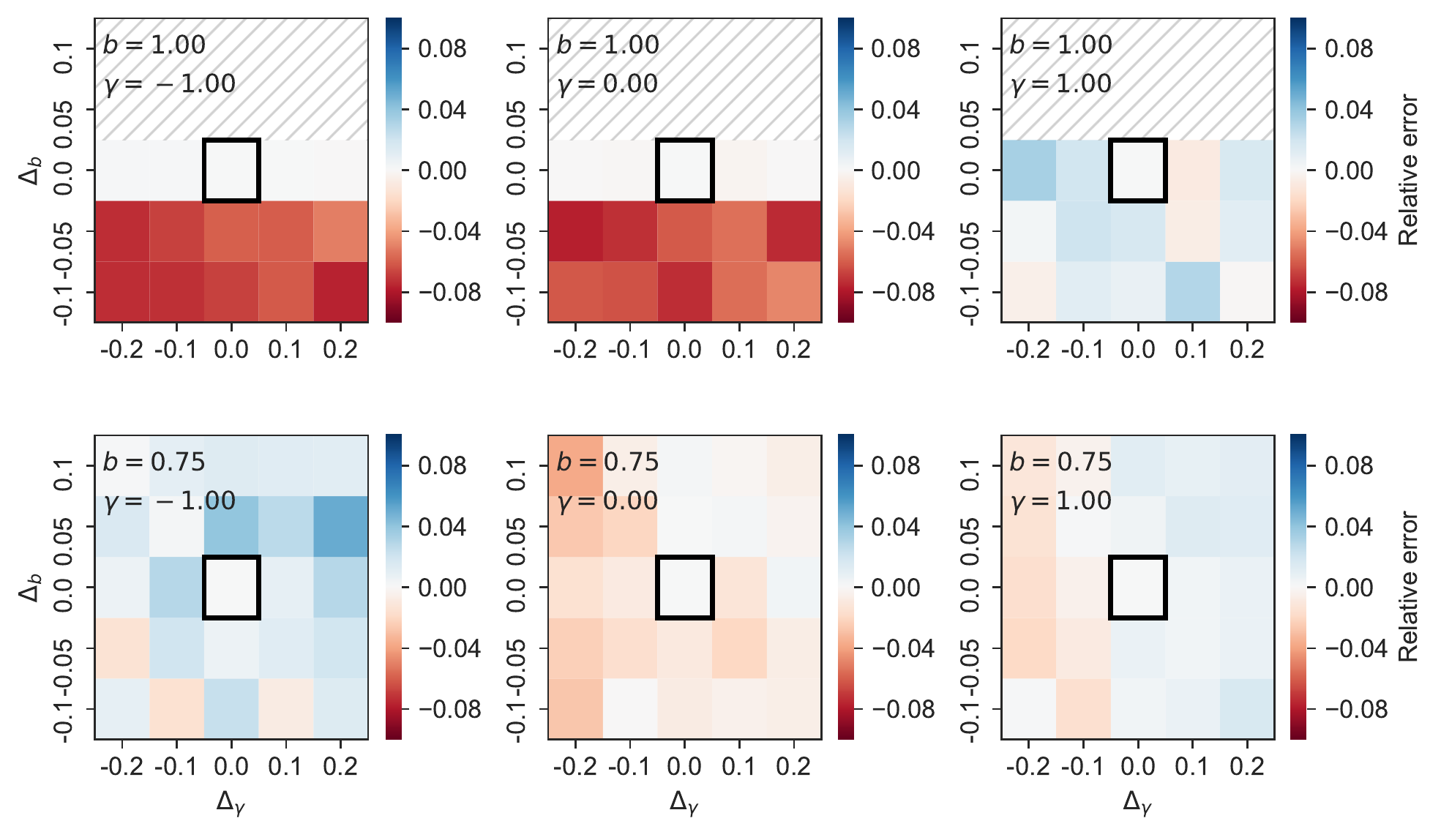}
    \caption[Sensitivity analysis of the parameter estimators]{
    \textbf{Sensitivity analysis of the parameter estimators.}
    Average correlation with the ground truth (top 6 panels) and relative error on the true correlation (bottom six panels) of the histories estimated by running the inference with misspecified parameters.
    These results are produced on trees ($b=1$, top rows) and on loopy graph ($b=0.75$, bottom rows), at various level of heterogeneity ($\gamma\in\{-1,0+1\}$ from left to right), also see inset text.
    The bold square shows the reference value (no perturbation) while the rest of the heat map shows results for absolute perturbations of magnitude $\Delta_\gamma, \Delta_b\in\{-0.2,-0.1,0.1,0.2\}$.
    Estimates above $\hat{b}=1$ are impossible and therefore not computed (shown as a hatched region).
    We use the sampling level most appropriate for each case  and $n=100\,000$ samples, and we average the results over 35 network instances.
    \label{fig:sensitivity}
    }
\end{figure}

The analysis, shown in Fig.~\ref{fig:sensitivity}, confirms that we can safely conduct temporal inference by conditioning on point estimates of the parameters.

\section{Maximum posterior maximization does not work}
\label{appendix:uniform}
In the main text, we mention that the posterior distribution $P(X_{0:T-1}|G,\gamma,b)$ is often uniform over large sets of histories, and that this epitomizes why posterior maximization (MAP) is not a suitable method to extract information from $G$.
We now demonstrate that the Krapivsky-Redner-Leyvraz generalization of preferential attachment \cite{krapivsky2000connectivity} is an extreme example of this problem.
Namely, we  explicitly show that when $\gamma\in\{0,1\}$, the posterior distribution of this model is uniform over all histories in $\Psi(G)$ 
\footnote{Lemma~5.3 of Ref.~\cite{magner2017times} states this fact without proof.}.
We then further argue that there exist large equivalence classes for general $\gamma\in\mathbb{R}$ and $b\in[0,1]$.

\subsection{Strict uniformity on trees}
Recall from Eqs.~\eqref{eq:general_growth_function} and \eqref{eq:model_kernel} that in the model where $b=1$, the logarithm of the unconditional probability of a history is given by
\begin{equation*}
    \log P(X_{0:T-1}|\gamma) = \sum_{t=1}^{T-1} \log u_{a_t}(\gamma, V_t),\; u_i(\gamma, V_t) = \frac{k_i^{\gamma}(t)}{\sum_{j\in V_t} k_j^{\gamma}(t)},
\end{equation*}
where $V_t$ is the node set of $G_t$ prior to any modification of the graph's structure, and
where we have denoted by  $a_t$ the node selected as the fixation site at time $t$ in $X_{0:T-1}$.
To demonstrate that the distribution is uniform over the set of all consistent histories, we first define the normalization $Z(t;\gamma)=\sum_{j\in V_t}k_j^{\gamma}(t)$ and rewrite
\begin{equation}
    \label{eq:prior_on_trees}
    \log P(X_{0:T-1}|\gamma) = \sum_{t=1}^{T-1}\left[\log k_{a_t}^{\gamma}(t) - \log Z(t;\gamma)\right]\;.
\end{equation}
Now, in the special cases of uniform attachment and linear preferential attachment, corresponding to $\gamma=0$ and $\gamma=1$, the normalization $Z(t;\gamma)$ \emph{always} takes a special value  independent from the actual content of $V_t$, namely
\begin{align*}
    Z(t;\gamma=0) &= \sum_{j\in V_t} k_j^0 = |V_t| = t + 1,\\
    Z(t;\gamma=1) &= \sum_{j\in V_t} k_j = 2t\;.
\end{align*}
The second identity follows from the fact that exactly one edge is created at each $t$, and that the sum of all degrees is always equal to twice the number of edges.
These normalizations are independent of $X_{0:T-1}$, meaning that they can be dropped as an additive constant.
Using $\gamma=0$ and $\gamma=1$ in  Eq.~\eqref{eq:prior_on_trees}, we are left with
\begin{equation}
     \log P(X_{0:T-1}|\gamma) \propto \left\{
     \begin{array}{ll}
        \mathrm{Constant}& \text{$\gamma=0$,}\\
        \sum_{t=1}^{T-1}\log k_{a_t}(t)& \text{$\gamma=1$.}
     \end{array}
     \;\right.
     \label{eq:special_prior}
\end{equation}
This last equation directly shows that the distribution $P(X_{0:T-1}|\gamma)$ (and therefore the posterior distribution) is uniform over all histories when $\gamma=0$.
Less obvious is the fact that the equation also implies a uniform posterior distribution in the case $\gamma=1$.
To see why  , notice that the posterior distribution is obtained by conditioning on $G$, and that this restricts the possible histories to those in which a node $i$ of degree $k^*_i$ in $G$ appears $k^*_i-1$ times in the sum $\sum_{t=1}^{T-1}\log k_{a_t}(t)$: once as a node of degree one, once as a node of degree two, etc.
Hence, every history consistent with $G$ is associated with some permutation of the same sum.
Obviously, a permutation does not change the value of the sum; therefore, the posterior distribution is uniform over all histories consistent with $G$.\\

\subsection{Extension to all parameters}
Large sets of equally likely histories also arise in the more general attachment model on trees (i.e., when $\gamma\in\mathbb{R}$ with $b=1$).
The proof that these sets of histories exist is similar in spirit to that of the special cases above.
We first make use of the permutation argument again, noting that it applies to the general sum $\sum_{t=1}^{T-1}\log k_{a_t}^{\gamma}(t)$, regardless of the value of $\gamma$.
The problem therefore reduces to the study of the evolution of the normalization constant.
Different from the special cases $\gamma=0$ and $\gamma=1$, the normalization $Z(t;\gamma)$ does not grow at the same rate for all histories when $\gamma$ is arbitrary.
But, as we now show, there are still equivalence classes with respect to the posterior distribution.
For example, consider two histories identical in all respects until a last node of degree $k$ and its $k-1$ remaining neighbors are encountered.
The $(k-1)!$ histories resulting from the enumeration of this neighborhood will have, by construction, equivalent sequences of normalization constants $Z(1;{\gamma})\to Z(2;{\gamma})\to\hdots Z(T-1;{\gamma})$, which imply that these histories will be associated with the same posterior probability, and that they will form a small equivalence class.
Broader equivalence classes can be identified by noticing that similar permutations arise not only at the end, but also at any point of the histories, and that they interact combinatorially: If there are $m$ such equivalent sets of edges, of sizes $k_1,...,k_m$, then each different point of the posterior is degenerated $k_1!\times ... \times k_m!$ times.
This argument trivially extends to any  $b\in[0,1]$.

\section{Optimality of the MMSE estimators}
\label{appendix:optimal}

The proof  that the posterior average of $\tau_X(e)$ maximizes correlation a posteriori goes as follows.

By a slight abuse of notation, let us refer to an estimated history constructed with some arbitrary estimators $\{\hat{\tau}(e)\}_{e\in E(G)}$  as $Y$, such that $\tau_Y(e)=\hat{\tau}(e)$.
Then, assuming again that $X$ is drawn from the posterior distribution, the expected correlation of $X$ and $Y$ is
\begin{align}
    \langle \rho(X,Y) \rangle
     = \sum_{X\in\Psi(G)} P(X|G,\gamma,b)\rho(X,Y),
\end{align}
which we rewrite as
\begin{align*}
    \frac{1}{\tilde{\sigma}_X\tilde{\sigma}_Y}  \sum_{X\in\Psi(G)} P(X|G,\gamma,b)\sum\limits_{e\in E(G)} \bigl(\tau_X(e) - \langle \tau \rangle\bigr)\bigl(\tau_Y(e) - \langle \tau \rangle\bigr)\notag\\
    =\frac{1}{\tilde{\sigma}_X\tilde{\sigma}_Y}  \sum\limits_{e\in E(G)} \Big[\langle \tau(e)\rangle - \langle \tau \rangle \Big]\Big[\tau_Y(e) - \langle \tau \rangle\Big],
\end{align*}
where $\tilde{\sigma}_X^2:=\sum_{e\in E(G)}(\tau_X(e)-\langle \tau \rangle )^2$ and $\tilde{\sigma}_Y^2$ are sums of individual variances (the same variances we use to calculate uncertainty $U(G)$, see Eqs.~\eqref{eq:variance} and \eqref{eq:uncertainy}).
We can to take these standard deviations out of the sum because (1) $\tilde{\sigma}_Y$ is independent of $X$, and (2) the value of $\tilde{\sigma}_X$ is constant for all $X$, since one edge must occupy each `time slot' $t=0,\hdots,T-1$ by definition, implying $\tilde{\sigma}_X^2 = \sum_{t=0}^{T-1} (t - \langle \tau\rangle)^2$.
Again, somewhat stretching the notation, we define $Z$ as the history constructed with the MMSE estimators, i.e, the history such that $\tau_Z(e)=\langle \tau(e)\rangle $.
We can then express the expected correlation as
\begin{equation}
    \label{eq:posterior_avg_correlation}
    \langle \rho(X,Y) \rangle  = \frac{\tilde{\sigma}_Z}{\tilde{\sigma}_{\!X}} \rho(Z,Y),
\end{equation}
where $\tilde{\sigma}_Z\neq \tilde{\sigma}_X$ in general.
Equation~\eqref{eq:posterior_avg_correlation} tells us that when histories are actually drawn from the posterior distribution,  the expected correlation of the arbitrary estimators $\{\tau_Y(e)\}$ is proportional to the correlation between these estimators and the MMSE estimators.
The expected overall correlation is therefore maximized if we choose $Y$ to be the MMSE estimators of the arrival times.

\section{Finding equivalent edges}
\label{appendix:orbits}
We gave an intuitive definition of equivalent edges in the main text.
A more formal definition can be given by using the concept of \emph{orbits}  \cite{mckay2014practical}.
An orbit is a set of nodes that map onto themselves when we take an automorphism of $G$.
Hence, for example, the three nodes of a graph comprising a single triangle form a single orbit.
As another example, the two end nodes ($a$ and $c$) of the graph with edges $\{(a,b),(b,c)\}$ are also form an orbit, while the central node $b$  forms an orbit by itself.

Our goal is, of course, to find the equivalent edges of $G$ and not the equivalent nodes, so we need to adapt the concept slightly.
One can easily define ``edge orbits''  in the same way, but standard graph isomorphism software tends to deal with node orbits only \cite{mckay2014practical}.
Instead, we take a shortcut and resort to the \emph{line graph} $L(G)$ of $G$.
The line graph $L(G)$ is constructed by replacing every edge of $G$ by a node, and by connecting two of these nodes if the corresponding edges share an endpoint in $G$.
This transformation is useful because the indistinguishable edges of $G$ are mapped to the (node) orbits of $L(G)$ by construction, since the $L(G)$ preserves the graph symmetries of $G$.
This method gives a straightforward algorithm to find equivalent edges: We transform $G$ into a line graph $L(G)$, run a standard orbit identification algorithm on $L(G)$, and translate the result back to $G$.

We note that the definition of the line graph is ambiguous when $G$ has self-loops and parallel edges.
(Should a self-loop be connected with itself in $L(G)$? What about sets parallel edges: Should the copies of an edge be connected with the other copies?)
Since our goal is to find the distinguishable edges of $G$,  we must ensure that the  transformation does not erase the important symmetries of $G$.
With this goal in mind, we choose to (i) \emph{not} connect the parallel edges of $G$ among themselves, and (ii) add a self-loop to the nodes of $L(G)$ that stand in for self-loops in $G$.
Choice (i) allows us to distinguish the  parallel edges from simple edges.
As an example, suppose we have a triangle with a parallel edge $\{(a,b),(a,b),(b,c),(c,a)\}$. 
The resulting line graph is a clique with a missing edge between the two copies of $(a,b)$, which allows us to disambiguate the parallel and simple edges.
Choice (ii) allows us to tell self-loops apart from other edges.
For example, if we have a star graph $G$ with added self-loops on the central nodes, then the line graph is again a clique, but the self-loops are marked as such---again allowing us to tell them apart from the spokes.

%

\end{document}